\documentclass[aps,prl,twocolumn,floatfix, superscriptaddress,footinbib]{revtex4-2} 

\usepackage{amsmath,amssymb}
\usepackage{braket}
\usepackage{mathtools}
\usepackage{xcolor, ulem}
\usepackage{siunitx}
\usepackage{CJK} 
\usepackage{lipsum}
\usepackage{graphicx}
\usepackage{footnotebackref}
\usepackage{pifont}

\usepackage{multirow}

\newcommand{\cmark}{\ding{51}}%
\newcommand{\xmark}{\ding{55}}

\definecolor{pal0}{rgb}{0.8941, 0.102 , 0.1098}
\definecolor{pal1}{rgb}{0.2157, 0.4941, 0.7216}
\definecolor{pal2}{rgb}{0.302 , 0.6863, 0.2902}
\definecolor{pal3}{rgb}{0.5961, 0.3059, 0.6392}
\definecolor{pal4}{rgb}{1.    , 0.498 , 0.    }

\newcommand*{\kekule}[0]{Kekul\'{e} }
\newcommand*{\Tp}[0]{\mathcal{T}'}
\newcommand*{\T}[0]{\mathcal{T}}
\renewcommand*{\v}[1]{\mathbf{#1}}

\DeclareMathOperator{\Tr}{Tr}

\usepackage{algorithm}
\usepackage[noend]{algpseudocode}
\usepackage{setspace}

\usepackage{pifont}

\begin{document}

    \begin{CJK*}{UTF8}{min}

	\title{Detecting  symmetry breaking in magic angle graphene using scanning tunneling microscopy}
	
    \author{Jung Pyo Hong}    \thanks{These authors contributed equally}
\affiliation{Department of Physics, Princeton University, Princeton, NJ 08540, USA}
    
    \author{Tomohiro Soejima (副島智大)}
    \thanks{These authors contributed equally}
\affiliation{Department of Physics, University of California, Berkeley, CA 94720, USA}
    
    \author{M. P. Zaletel}
\affiliation{Department of Physics, University of California, Berkeley, CA 94720, USA}
\affiliation{Materials Sciences Division, Lawrence Berkeley National Laboratory, Berkeley, CA 94720, USA}
	\date{\today}
	
	\begin{abstract}


A growing body of experimental work suggests that magic angle twisted bilayer graphene  exhibits a  ``cascade''  of spontaneous symmetry breaking transitions, 
 sparking interest in the potential relationship between symmetry-breaking and superconductivity. However, it has proven difficult to find experimental probes which can unambiguously identify the nature of the symmetry breaking. Here we show how atomically-resolved scanning tunneling microscopy can be used as a fingerprint of symmetry breaking order.
 By analyzing the pattern of sublattice polarization and ``Kekul\'{e}'' distortions in small magnetic fields, order parameters for each of the most competitive symmetry-breaking states can be identified.
 In particular, we show that the ``Kramers intervalley coherent state,'' which theoretical work predicts to be the ground state at even integer fillings, shows a Kekul\'{e} distortion which emerges only in  a magnetic field. 

	\end{abstract}
	\maketitle
	
	\end{CJK*}

Since superconductivity and correlated insulators were found in magic angle twisted bilayer graphene (MATBG) \cite{cao2018Correlated, cao2018Unconventional, yankowitz2019Tuning}, there has been a vigorous attempt to identify the nature of the ground states at integer fillings. 
While widely believed to arise from spontaneous symmetry-breaking in the space of spin and valley, theoretical works have identified a wealth of candidate states which are close in energy, including insulators in the $U(4)\times U(4)$ manifold \cite{bultinck2020Ground, vafek2020Hidden, lian2021TBG}, a nematic semi-metal (nSM) \cite{choi2019Electronic,liu2021Nematic}, and the incommensurate \kekule spiral (IKS) \cite{kwan2021Kekul}. However, while transport and thermodynamic quantities  can be used to establish the insulating vs. metallic nature of the states,  clear experimental identification of the ground state order has proven difficult.

\begin{center}
\begin{figure*}
    \includegraphics[width=1.0\textwidth,height=0.5\textheight,keepaspectratio]{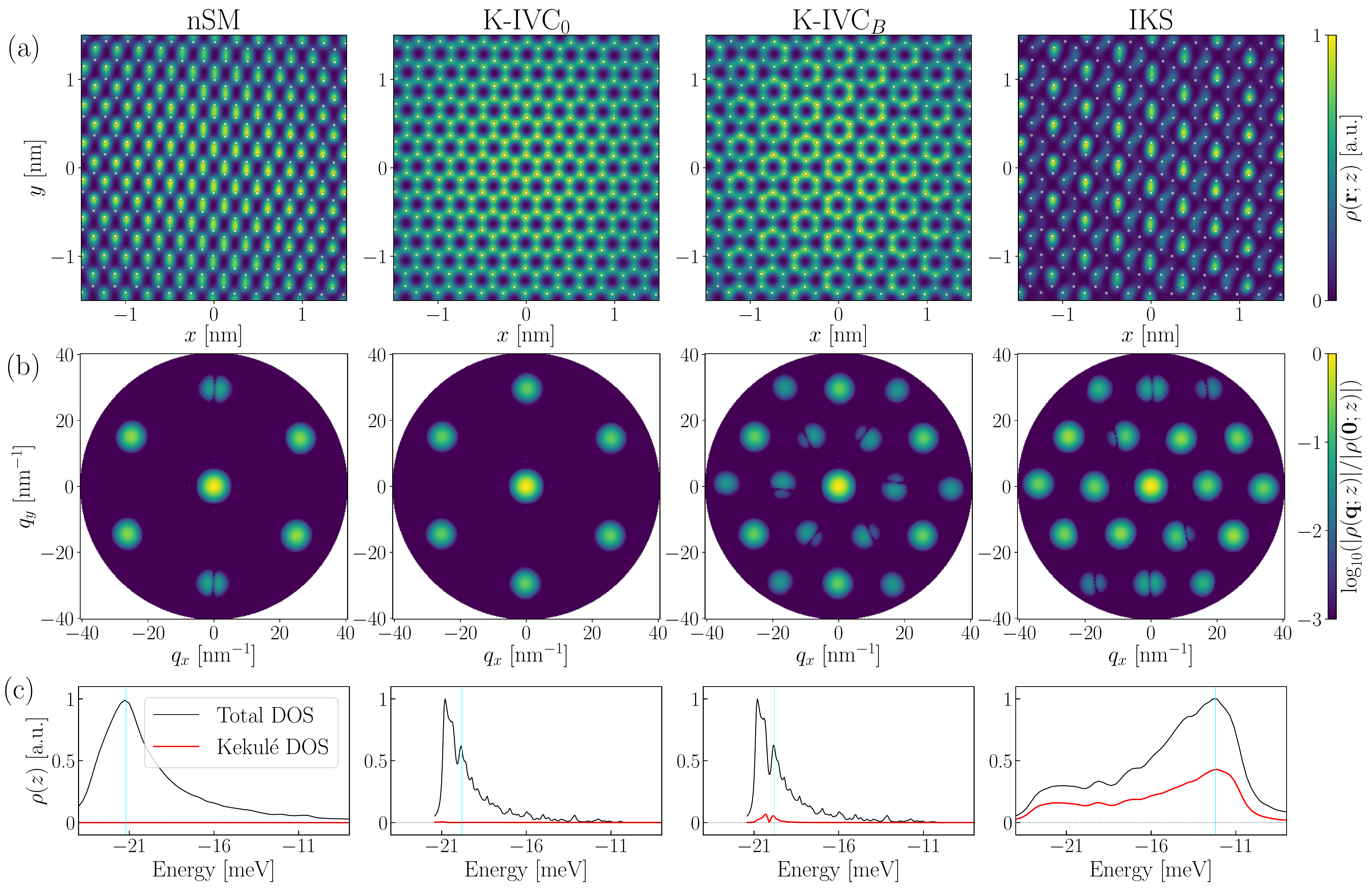}
    \caption{Spectroscopy calculation results for the self-consistent HF solutions of spinless nSM, $\text{K-IVC}_{0}$, $\text{K-IVC}_{B}$ states at CNP ($\nu=0$) and IKS states at $\nu=-1$. We present (a) total LDOS $\rho(\v{r};z)$ (b) total FTLDOS $\rho(\v{q};z)$ (c) total and \kekule-DOS $\rho(z=\omega+i\eta)$ signals for the bottom layer AA-region of MATBG. The signals are normalized by their maximum. For FTLDOS, logarithm is taken on the normalized signals to highlight relative magnitudes between the peaks. For K-IVC$_{0}$ and K-IVC$_{B}$ we set the Lorentzian broadening parameter $\eta=\SI{0.1}{\milli\electronvolt}$, while for nSM and IKS $\eta=\SI{0.5}{\milli\electronvolt}$. The scanning-energies $\omega=-21.2$, $-19.8$, $-19.8$, \SI{-12.2}{\milli\electronvolt} (cyan lines in (c)) are chosen respectively for each phase to compute LDOS and FTLDOS. Energies below \SI{-8}{\milli\electronvolt} are presented in (c) to show DOS signals for the occupied states. White dots in (a) denote carbon lattice sites.}
    \label{fig:ldos_ftldos_dos}
\end{figure*}
\end{center}

Scanning tunneling microscopy (STM), which measures the local density of states (LDOS), and by spatial integration the DOS itself, is a promising tool for distinguishing between these phases. Thus far STM measurements in MATBG have  largely  focused on  modulations in the LDOS at the Moir\'{e}-scale \cite{kerelsky2019Maximized, wong2020Cascade, xie2019Spectroscopic, choi2019Electronic, jiang2019Charge}.
STM has found evidence for a ``cascade'' of putative symmetry breaking transitions \cite{wong2020Cascade, choi2021Correlationdriven}, but has not yet detected the relevant order parameters.
Take, for example, the Kramers-intervalley coherent (K-IVC) state \cite{bultinck2020Ground,zhang2020Correlated, lian2021TBG, kwan2021Kekul} and the nSM \cite{liu2021Nematic}, which are believed to compete at the charge neutrality point (CNP) \cite{parker2021StrainInduced, hofmann2021Fermionic, pan2021Dynamic}. While there are quantitative differences between the DOS of the two phases (they are gapped and semimetallic respectively),  given finite energy resolution their DOS are in practice very similar (Fig.~\ref{fig:ldos_ftldos_dos}(c)), making  discrimination difficult \cite{parker2021StrainInduced}. Furthermore, the Moir\'{e}-scale spatial modulation of the LDOS is largely dominated by the peaks at the triangular-lattice of AA-regions \cite{kerelsky2019Maximized, jiang2019Charge, choi2019Electronic}. As this feature arises from the electronic structure of the flat bands, it is  common to all the phases,  making it inadequate for identifying the ground state.
What is needed is a probe capable of directly imaging the order parameter of the symmetry breaking, e.g. graphene-scale translation symmetry breaking and nematicity for the K-IVC and nSM respectively. 

In this work we show that atomically-resolved STM measurements  are an ideal method for distinguishing between these competing states.
The K-IVC can be detected by the formation of a $\sqrt{3} \times \sqrt{3}$ \kekule pattern \cite{gutierrez2016Imaginga, li2019Scanninga, ma2018Modulating, fiori2017Liintercalated, shimizu2015ChargeDensity, kanetani2012Ca, liu2021Visualizing} only in the presence of a small ($B \sim \SI{1}{T}$) magnetic field, while other states can be detected via  sublattice polarization and bond nematicity.
Combined with transport measurements, the following  states can be distinguished: 1) symmetric Dirac semimetal 2) nSM 3)  K-IVC 4) Generic IVC state (e.g. IKS) 5) valley Hall and 6) valley polarized (Table.~\ref{tab:phases}, Fig.~\ref{fig:ldos_ftldos_dos}).

The ability of atomically-resolved STM to detect symmetry breaking in low-density flat-bands was recently demonstrated experimentally by Liu et al.~\cite{liu2021Visualizing}.
In the zeroth Landau-level of un-twisted monolayer graphene,  Coulomb interactions favor the formation of competing quantum Hall ferromagnets including valley-polarized, Kekul\'{e}, and canted-antiferromagnetic orders.
By imaging the atomically-resolved DOS, Liu et al. were able to directly measure the order parameters of the valley-polarized and \kekule states. 
As the Landau-levels and MATBG bands have comparable electron densities and energy scales, their findings are encouraging for the analogous measurements in MATBG.

\begin{figure}
    \centering
    \includegraphics[width=\linewidth]{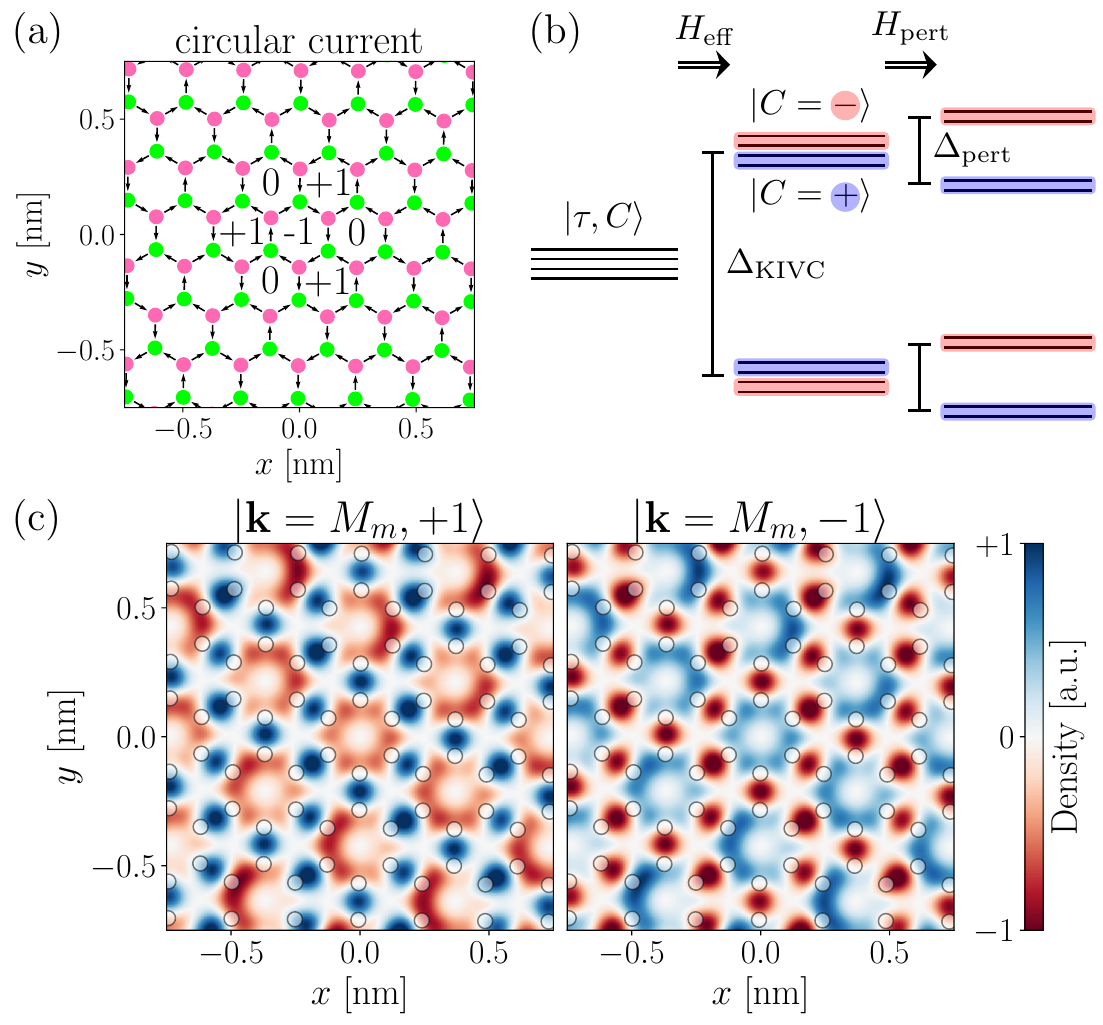}
    \caption{(a) Total current-density of K-IVC eigenstates (Eq.\eqref{eq:density_matrix_basis}) $\ket{\v{k}=M_{m}, n=\pm 1}$ for each plaquette. $M_{m}$ is the $M$-point of the mini-BZ. Lime/pink dots signify $A/B$ sublattice sites. It features a $\sqrt{3}\times\sqrt{3}$ \kekule pattern of circulating currents. (b) Schematic representation of the energy levels of K-IVC. From left to right: the basis states; K-IVC eigenstates; Perturbation eigenstates (Eq.\eqref{eq:H_pert_momentum_space}). Perturbation breaks the Kramer's degeneracy of K-IVC eigenstates. Red(blue) shading corresponds to superpositions with predominantly Chern $+1$ ($-1$) character.
    (c) \kekule charge-density of $\ket{\v{k}=M_{m},\text{occ},n=\pm 1}$. They have opposite \kekule signals due to $\Tp$ symmetry and their sum cancels out.}
    \label{fig:mechanism}
\end{figure}

The fact that the K-IVC can be distinguished from a generic IVC state may come as a surprise. Indeed, generic IVC states, which have coherence between two valleys, are always expected to show a \kekule pattern which triples the unit cell at the scale of the graphene lattice.  However, the K-IVC features an anti-unitary ``Kramer's time-reversal''  symmetry $\Tp$  which  extinguishes the charge density wave (CDW); instead there is a  magnetization density wave (MDW), e.g., a \kekule pattern of circulating currents \cite{bultinck2020Ground} (Fig.~\ref{fig:mechanism}).
However, breaking the $\Tp$ symmetry with a small magnetic field converts the MDW into an CDW, a unique fingerprint of the K-IVC which is observable in STM spectroscopy.
Taken together, the proposed experiment is a promising method for identifying the nature of the ground state without resorting to quantitative details.

\textit{Active bands of  twisted bilayer graphene.} We first review the flat band physics of MATBG and and set  notation. We focus on a single spin-species for simplicity, and will return to the question of spin structure later. 
A convenient basis for the four flatbands (per spin) of MATBG is the ``Chern'' (or chiral) basis $\ket{\v{k}, \tau, C}$, where $\v{k}$ is crystal momentum in the mini-BZ, $\tau = \pm 1$ is the valley label, and $C = \pm 1$ is the Chern number \cite{tarnopolsky2019Origin, bultinck2020Ground}.
To good approximation, the chiral basis is sublattice-polarized according to $A/B = \sigma = C \tau$.
We  fix  two-fold rotation $\mathcal{C}_2$ and spin-less time-reversal $\T$ symmetry to act as
\begin{equation}
    \mathcal{C}_2 \ket{k, \tau, C} = \ket{-\v{k}, -\tau, C}, \mathcal{T} \ket{\v{k}, \tau, C} = \ket{-\v{k}, -\tau, -C}.
\end{equation}
Neglecting small umklapp terms, the Hamiltonian is invariant under valley-dependent phase rotation $U_V(1)$. Using the Pauli matrix notation for $\tau$, the action of $U_V(1)$ is written as $e^{i\tau_z \theta} \ket{\v{k}, \tau, C} = e^{i\tau \theta}\ket{\v{k}, \tau, C}$.

\textit{STM-spectroscopy of the competing phases.}  We first numerically compute the LDOS of MATBG
starting from self-consistent Hartree-Fock (HF) ansatzes for the various competing phases.
We consider a Hamiltonian in which the Coulomb interaction is projected into the Bistritzer-MacDonald (BM)  continuum model \cite{bistritzer2011Moire}.
While most of our findings follow from general symmetry considerations, several studies have shown that HF is accurate at even-integer fillings \cite{xie2020Nature, bultinck2020Ground, liu2021Theories, cea2020Band, zhang2020Correlated, lian2021TBG}, in some cases producing ground states nearly identical to the semi-exact solutions obtained from DMRG and exact diagonalization \cite{kang2020NonAbelian, soejima2020Efficient, parker2021StrainInduced, xie2021Twisted}.
The single-particle spectrum $\ket{E_i}$ of the self-consistent Hartree-Fock Hamiltonian $H_{\textrm{eff}}$ is then used to obtain the LDOS according to
\begin{align}
    \textrm{LDOS}(\omega = e V, \mathbf{r})  &= \frac{1}{\pi} \Im \sum_i \frac{\left |\braket{\mathbf{r} | E_i}\right|^2}{E_i - \omega + i \eta}
\end{align}
where $\omega$ is the bias voltage of STM, $\eta$ 
is a phenomenological Lorentzian-broadening parameter.
In order to obtain the continuum LDOS at $\mathbf{r}$, we assume the Wannier orbitals $w_{A/B}(\mathbf{R})$ of the carbon atoms take an approximately Gaussian form. While they are orthogonal, they do have some spatial overlap, which causes coherences between neighboring sites to modulate the density on the honeycomb bonds, a crucial ingredient for producing the observed patterns.
We may similarly obtain the Fourier-transformed LDOS $(\text{FTLDOS})\,\rho(\v{q};\omega)$, and the total $\text{DOS}\,\rho(\omega)$.
We refer to the Supplementary Material \footnote{See Supplementary Material
for details of Green's function formalism, Hartree-Fock calculation, and other numerical data} for the full specification of the model, Hartree-Fock calculations, and LDOS calculations.

Our numerical calculations focus on the states shown by prior work to be most competitive when filling 2 of 4 bands: the nSM \cite{liu2021Nematic, soejima2020Efficient, kang2020NonAbelian}, K-IVC \cite{bultinck2020Ground, lian2021TBG, zhang2020Correlated,  parker2021StrainInduced}, and IKS \cite{kwan2021Kekul}. Their properties are summarized in Table.~\ref{tab:phases}.
The K-IVC qualitatively changes in a perpendicular magnetic field, so we let K-IVC$_{0/B}$ indicate the absence/presence of a magnetic field.
The valley-Hall (VH), valley-polarized (VP) and symmetric Dirac semi-metal (DSM) states will be discussed in-text.

All states were obtained from HF using the methods described in  \cite{bultinck2020Ground, Note1}.
In order to obtain the nSM and IKS as self-consistent solutions, a small heterostrain ($0.04 \%$ for nSM, $0.2 \%$ for IKS) in the $x$-direction was added to stabilize them \cite{Note1}. 
To produce the K-IVC$_{B}$ phase we focus on the effect of the orbital Zeeman splitting produced by the $B$-field, which we model via the phenomenological Hamiltonian $H_\text{pert} = E_B C$, where $C$ is the Chern number of the band and $E_B = \SI{0.1}{\milli\electronvolt}$, as will be discussed in a subsequent section.

In Fig.~\ref{fig:ldos_ftldos_dos}, we present the total LDOS/FTLDOS/DOS of the bottom layer in the region of AA-stacking, where most of the DOS resides (the AB/BA-regions are also interesting; see SM\cite{Note1}). We now walk through the features which are characteristic of each phase.

(i) Symmetric Dirac semi-metal (DSM): the ground state of the non-interacting BM model \cite{bistritzer2011Moire} (or a dressed version thereof) respects all symmetries of MATBG. In particular, it has $\mathcal{C}_3$ and $U_V(1)$ symmetry, which prohibits a nematic axis and \kekule signal, respectively.

(ii) nSM: The LDOS is nematic,  strongly breaking $\mathcal{C}_{3}$ but preserving $\mathcal{C}_{2}$.
The order manifests in two forms: first, in the orientation of the strong bonds (the magnitude of this effect is sensitive to choice of carbon Wannier orbital), and second, while the very center of the AA-region has  equal weight on the two sublattices, to the right(left) the dominant weight shifts to the $A$($B$) sublattice. This polarization is not observed along the axis rotated by $2 \pi / 3$.
The orientation and quantitative magnitude of  both features are sensitive to the (weak) applied strain.
These features also manifest in the FTLDOS, where  \emph{two} of the six Bragg peaks (at the reciprocal vectors of the graphene lattice) have a nodal line across which the phase changes over the mini-BZ.
Similar nematic behavior was observed in earlier STM experiments \cite{jiang2019Charge}, as we discuss later.

(iii-a) $\text{K-IVC}_{0}$: The LDOS respects all the symmetries.
The absence of a \kekule pattern despite the inter-valley coherence is  enforced by the $\Tp$-selection rule to be derived later. The FTLDOS is peaked only at the graphene reciprocal vectors.

(iii-b) $\text{K-IVC}_{B}$: For \emph{a small range of tunneling bias} the LDOS  shows an atomic-scale distortion which forms a $\sqrt{3}\times\sqrt{3}$ \kekule pattern respecting $\mathcal{C}_{3}$. The FTLDOS exhibits dominant peaks at the Bragg points and subdominant (of order $10^{-1}$) peaks at intervalley-scattering momenta $K - K'$. 
We emphasize that the \kekule signal only shows up for a range of bias close to the van-Hove peak (see \kekule-DOS, Fig.~\ref{fig:ldos_ftldos_dos}(c)), and in fact the
signal changes sign as the peak is crossed.  The origin of this energy dependence will become clear shortly. 

(iv) IKS: The LDOS exhibits a $\sqrt{3}\times\sqrt{3}$ distortion which breaks $\mathcal{C}_3$
\footnote{IKS (and likewise K-IVC) state enjoys $\mathcal{C}_2\mathcal{T}$ symmetry only when $\theta_\text{IVC}$ takes particular values}. 
The \kekule-DOS is on the same order as the total DOS over the entire range of occupied band energies \footnote{ As demonstrated in \cite{Note1}, the \kekule-DOS signals however vanish for some of the conduction bands due to loss of inter-valley coherence.}.

(v) VH: The valley-Hall state is obtained by occupying  either the $\sigma = A$ or $B$ sublattice (e.g. $\ket{\tau = 1, C=1}, \ket{\tau = -1, C=-1}$), breaking $\mathcal{C}_2$. We thus expect strong sublattice polarization, with opposite sign in the filled / empty DOS.

(vi) VP: The valley-polarized state is obtained by doubly occupying one valley (e.g. $\ket{\tau = 1, C=1}, \ket{\tau = 1, C=-1}$), breaking $\mathcal{C}_2$ and $\mathcal{T}$ while preserving $\mathcal{C}_3, \mathcal{C}_2 \mathcal{T}$. $\mathcal{C}_2 \mathcal{T}$ rules out sublattice polarization, so the LDOS is fully symmetric. 
However, when $B > 0$, $\mathcal{C}_2 \mathcal{T}$ is broken and we expect sublattice polarization to emerge in a narrow range of tunneling voltages (c.f. our discussion of the K-IVC$_B$ phase.)

We note that while we have discussed the signatures in real space, experimentally it is most convenient to extract the order parameters from the phase structure of the FTLDOS, as described in Ref.~\cite{liu2021Visualizing}.
This procedure gives a precise determination of the sublattice polarization and the phase of the inter-valley coherence $\theta_{IVC}$, as reviewed in the SM\cite{Note1}.

\begin{table}[]
    \centering
\begin{tabular}{c|cccccccc}
\multirow{2}{*}{Phase} &\multirow{2}{*}{$U_V(1)$} & \multirow{2}{*}{$\Tp$} & \multirow{2}{*}{$\mathcal{C}_3$} & \multicolumn{2}{c}{\kekule} & \multicolumn{2}{c}{S.L. Pol.} \\
 & &  &  & $B=0$  & $B\neq 0$ &   $B=0$  & $B\neq 0$  \\ \hline
DSM   &     \cmark     &  \cmark     & \cmark &  \xmark & \xmark & \xmark & \xmark \\
nSM   &     \cmark     &  \cmark     & \xmark & \xmark & \xmark & \xmark & \xmark \\
VH   &     \cmark     &  \cmark     & \cmark & \xmark & \xmark &  \cmark   &  \cmark  \\
VP   &     \cmark     &  \xmark     & \cmark & \xmark & \xmark &   \xmark   &  \cmark   \\
\hline
K-IVC  & \xmark &    \cmark  & \cmark & \xmark & \cmark &  \xmark & \xmark  \\
IKS   &      \xmark    &   \xmark & \xmark   & \cmark & \cmark &  \xmark & \xmark  & 

\end{tabular}
    \caption{Symmetry properties of various ground state candidate states. DSM stands for symmetric Dirac semimetal. The column for \kekule denotes absence (\xmark) or presence (\cmark) of \kekule pattern in LDOS. The column for S.L. Pol. indicates whether the LDOS is $A/B$ sublattice polarized at the very center of the AA-stacking region, when averaging over $3$ unit cells to remove contributions from the \kekule signal. }
    \label{tab:phases}
\end{table}

\textit{Vanishing \kekule signal of the K-IVC state.}
We now explain  why the K-IVC \kekule signal emerges only in a magnetic field. 
In Ref.~\cite{bultinck2020Ground} it was shown that the K-IVC state produces a \kekule-like pattern of circulating \emph{currents} (e.g. a magnetization density wave), not charge, as shown in Fig.~\ref{fig:mechanism}(a).
Since STM is sensitive to charge rather than current, it is intuitive that this order will escape notice. 
We may formalize this extinction as a selection rule.
While the K-IVC breaks $\mathcal{T}$, it preserves a modified ``Kramer's'' time-reversal $\mathcal{T}' = \tau_z\mathcal{T}$ which applies a $\pi$-phase rotation between the valleys \cite{bultinck2020Ground}.
To assess its consequences, we analyze the transformation properties of the FTLDOS
\begin{equation} \label{eq:FTLDOS_1}
    \rho(\v{q};z = \omega + i \eta) = \frac{-1}{2\pi i }  \left( \Tr[\hat{\rho}_{\v{q}} \hat{G}(z) ]  - \overline{\Tr[\hat{\rho}_{-\v{q}} \hat{G}(z) ]} \right )
\end{equation}
where $\hat{G}(z)$ is the electron Green's function
and $\hat{\rho}_{\v{q}}=e^{-i\v{q}\cdot\hat{\v{r}}}$ is the density operator \cite{Note1}. 
The ``\kekule-LDOS'' is the portion of the FTLDOS at inter-valley momentum transfer $\v{q} = K - K' + \Delta \v{q}$, where  $\Delta \v{q}$ is small. 

To derive selection rules for the \kekule-LDOS we consider either a unitary symmetry with $\mathcal{U}^{-1} \hat{G}(z) \mathcal{U}= \hat{G}(z)$ or an anti-unitary symmetry with $\mathcal{K}^{-1} \hat{G}(z) \mathcal{K}= \hat{G}^\dagger(z)$.
Suppose further that for some $\mathbf{q}$ of interest the density transforms either as $\mathcal{U}^{-1} \hat{\rho}_\v{q} \mathcal{U} = \pm \hat{\rho}_{\v{q}}$, or $\mathcal{K}^{-1} \hat{\rho}_\v{q} \mathcal{K} = \pm \hat{\rho}_{-\v{q}}$, where the sign $\pm$ will depend on the symmetry. 
By inserting these transformations into Eq.\eqref{eq:FTLDOS_1} \cite{Note1}, we 
obtain $\rho(\v{q};z) = \pm \rho(\v{q};z)$.
The odd case then enforces an extinction. 

For the case at hand, we expand the inter-valley part of $\hat{\rho}_{\v{q}}$ in a plane-wave basis $\braket{\v{r} | \v{k}, \tau} \propto e^{i(\v{k}+K_{\tau})\cdot\v{r}}$
\begin{equation}
    \hat{\rho}_{\v{q}} = \sum_\v{k} \ket{\v{k}+\Delta\v{q}, \tau = 1} \bra{\v{k}, \tau = -1},
\end{equation}
where $\v{k}$ is restricted to the vicinity of the graphene Dirac points. 
In the absence of IVC order, the symmetry $\mathcal{U} = \tau_z \in U_V(1)$ gives $\mathcal{U}^{-1} \hat{\rho}_\v{q} \mathcal{U} = - \hat{\rho}_{\v{q}}$, enforcing an extinction, while for a generic IVC (e.g. the absence of other symmetries) the \kekule-LDOS will be present. 
For the K-IVC we instead leverage $\Tp  \ket{\v{k}, \tau} = \tau \ket{-\v{k}, -\tau}$, giving
$\mathcal{\Tp}^{-1} \hat{\rho}_\v{q} \mathcal{\Tp} = - \hat{\rho}_{-\v{q}}$, and  conclude the K-IVC has vanishing \kekule-LDOS.

\textit{Effect of a $\Tp$-breaking perturbation.---} 
In the presence of a $\Tp$-breaking perturbation $H_{\text{pert}}$ - e.g., an applied perpendicular magnetic field $B$ - the selection rule is inoperative and the K-IVC$_B$ phase will generically manifest a \kekule pattern.
The emergence of a \kekule pattern at $B > 0$ is the smoking gun signature of K-IVC order.
In  order to understand its magnitude and the sensitive $E$-dependence found in Fig.~\ref{fig:ldos_ftldos_dos}(c), we now analyze an approximate form of $H_{\text{pert}}$ in detail.
 
An out-of plane $B$-field will have two effects. First, it will reconstruct the flat bands into a Hofstadter butterfly; however this effect is small for weak ($B < \SI{1}{T}$) fields, where $e B L_M^2 \ll \hbar$. Second, $B$ will couple to the orbital magnetic moment  $m(\v{k}, \tau, C) = \mu_B g(\v{k}, \tau, C)$ of the flat bands,  where $g \sim 2 -10$ \cite{bultinck2020Mechanism, zhu2020VoltageControlled, tschirhart2021Imaging}.
For simplicity we neglect the $\v{k}$-dependence, in which case symmetry enforces the simpler form $m(\v{k}, \tau, C) =  \mu_B g \,C$, so that $H_\text{pert} = E_B C$. For a $B = \SI{1}{\tesla}$ field, $E_B \sim  \SI{0.1}{\milli\electronvolt}$ is thus a conservative estimate of its magnitude.

\begin{figure}
    \centering
    \includegraphics[width=\linewidth]{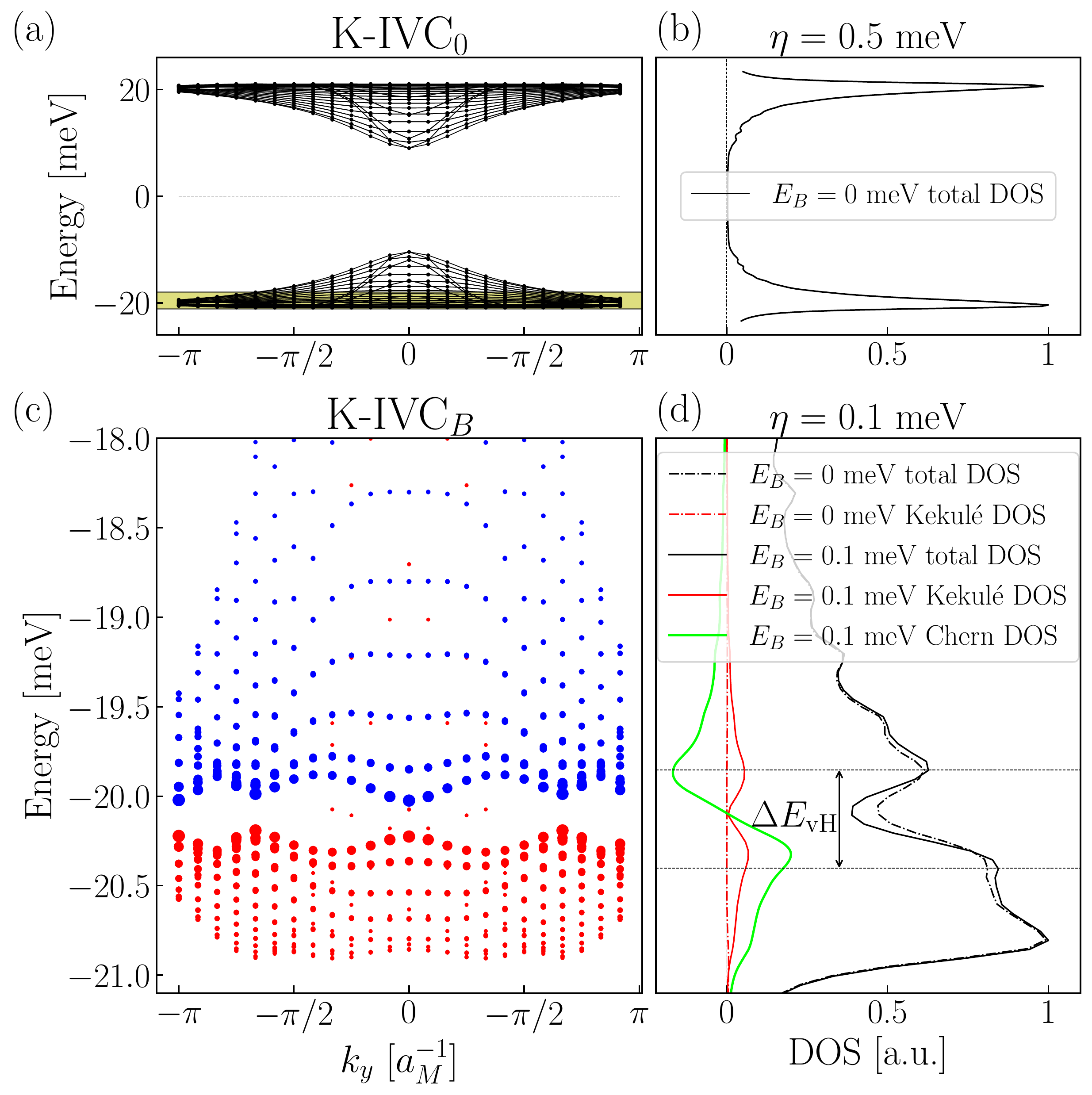}
    \caption{ (a-b) Self-consistent HF band-structure and DOS of K-IVC$_{0}$ at CNP and $\eta =$ \SI{0.5}{\milli\electronvolt}. Yellow patch denotes the spectrum within valence bands probed at $\eta =$ \SI{0.1}{\milli\electronvolt} in (c-d). (c) Band-structure of K-IVC$_{B}$ at $E_B = 0.1$ meV. Area of the blue (red) dots corresponds to the degree of positive (negative) Chern polarization of the wavefunction. (d) Total/\kekule-DOS of $\text{K-IVC}_{0}$ and $\text{K-IVC}_{B}$ along with Chern polarization-weighted total DOS. Splitting of the van-Hove singularity peaks for $\text{K-IVC}_{0}$ is estimated $\Delta E_{\text{vH}}\sim$ \SI{0.5}{\milli\electronvolt}.}
    \label{fig:chern_polarized_dos}
\end{figure}

To compute the change in the LDOS for energies below the Fermi level, we diagonalize $H_\text{eff} + H_\text{pert}$, as shown schematically in Fig.~\ref{fig:mechanism}(b).
Since $H_{\text{pert}}$ is small compared to the gap $\Delta_{\text{KIVC}}  \sim 20 - 40$ \si{\milli\electronvolt} of $H_\text{eff}$ (Fig.~\ref{fig:chern_polarized_dos}(a)), we project $H_\text{pert}$ into the space spanned by the two occupied eigenstates $|\v{k},n = 0/1\rangle$  of $H_\text{eff}$.
To constrain the form of $H_\text{pert}$, we combine $\mathcal{C}_2$, $\mathcal{T}$, and a relative valley phase to obtain a second symmetry of the K-IVC, $\mathcal{C}_2  \mathcal{T}'' = \mathcal{C}_2  \mathcal{T} e^{i\tau_z (\theta_\text{IVC} - \pi/2)/2}$, which acts locally in $\v{k}$.
Because $E_B C$ anti-commutes with  $\mathcal{C}_2  \mathcal{T}''$, the projection is constrained to take the general form
\begin{equation}
    \label{eq:H_pert_momentum_space}
    [H_\text{eff} + H_\text{pert}](\v{k}) = 
    \begin{pmatrix}
    E_0(\v{k}) & 0 \\
    0 & E_1(\v{k}) \\
    \end{pmatrix}
    +
    \begin{pmatrix}
    0 & \Delta_B(\v{k}) \\
    \overline{\Delta}_B(\v{k}) & 0 \\
    \end{pmatrix}.
\end{equation}
The effect of the perturbation is controlled by the ratio of $\Delta E (\v{k}) = E_1 (\v{k}) - E_0 (\v{k})$ and the magnetic perturbation $\Delta_B(\v{k})$. In Fig.~\ref{fig:chern_polarized_dos}(a,c) we see that $\Delta E(\v{k})$ is much smaller than the bandgap $\Delta_{\text{KIVC}}$ across most of the mini-BZ, which will thus result in a significant \kekule response in the LDOS even while the ground-state itself  changes by a negligible amount of order $H_{\text{pert}} / \Delta_{\text{KIVC}} \ll 1$.
For this reason, the energy resolution of STM is  crucial for observing an effect.

It is instructive to construct eigenstates of $H_\text{pert}$ in terms of the original valley/Chern basis $\ket{\v{k}, \tau, C}$.
In the strong-coupling limit, the  occupied K-IVC  states are spanned by the two states \cite{bultinck2020Ground}
\begin{align} \label{eq:density_matrix_basis}
    \ket{\v{k}, \textrm{occ}, +} & \approx (\ket{\v{k}, +, +} + e^{i\theta_\text{IVC}}\ket{\v{k}, -, +} )/\sqrt{2} \\
    \ket{\v{k}, \textrm{occ}, -} & \approx (\ket{\v{k}, +, -} - e^{i\theta_\text{IVC}}\ket{\v{k}, -, -} ) / \sqrt{2}
\end{align}
with definite Chern number $C = \pm 1$.
This is precisely the basis which diagonalizes $H_\text{pert} = E_B C$, while the K-IVC eigenstates $|\v{k},n\rangle$ are a linear-combination thereof.
Individually, each Chern sector $\ket{\v{k}, \textrm{occ}, \pm}$ contributes a \kekule density to the LDOS, as shown in Fig.~\ref{fig:mechanism}(c). However, $\mathcal{T}'$ ensures that the \kekule contribution from  $\ket{\v{k}, \textrm{occ}, +}$ and $\ket{-\v{k}, \textrm{occ}, -}$ cancel \cite{Note1}. 
But once split by $E_B C$, their contributions to the LDOS are shifted in energy, and a net signal appears.

\textit{K-IVC band-structure.---} 
As an explicit illustration of the $\Tp$-breaking mechanism we compare the band-structures of K-IVC$_{0}$ and K-IVC$_{B}$. The occupied DOS of KIVC$_{0}$  (Fig.~\ref{fig:chern_polarized_dos}(d)) has a dominant peak at $E_{\text{vH}} \sim \SI{-20}{meV}$ which is in fact composed of \emph{two} van-Hove singularities separated by $\Delta E_{\text{vH}}\sim$ \SI{0.5}{\milli\electronvolt}. These two peaks originate from two different bands $\ket{\v{k}, n = 0/1}$, so we may estimate $|\Delta E(\v{k})|\sim\Delta E_{\text{vH}}$. We thus expect substantial splitting of the Chern sectors whenever $|E_{B}| \sim \Delta E_{\text{vH}}$.
Indeed, the eigenstates of K-IVC$_{B}$ shows substantial Chern-polarization (Fig.~\ref{fig:chern_polarized_dos}(c)) for $E_{B} =$ \SI{0.1}{\milli\electronvolt}.
When probing the LDOS at energies slightly above / below $E_{\text{vH}}$, we couple predominantly to either the $C=1$ or $-1$ sector, and hence their \kekule signals (Fig.~\ref{fig:mechanism}(c)) no longer cancel.
We thus predict that the amplitude of the \kekule  signal will change sign as the tunneling bias  is swept across $E_{\text{vH}}$; for the experiment to work, it is thus crucial that the broadening $\eta$ remain smaller than the $B$-induced splitting $E_B$. We note that $E_B=\SI{0.1}{\milli\electronvolt}$ $\sim$ \SI{1}{\kelvin}, which is well-within experimental resolution.

\noindent\textit{Spin structure at $|\nu| = 0, 2$.---}
Finally, we reintroduce the spin degree of freedom $s^\mu$.
Most theoretical models discussed in the literature have an enhanced $SU(2)_+ \times SU(2)_-$ symmetry in which spins can be rotated independently within each valley.
The sign of the small ``Hund's'' coupling  which breaks this symmetry down to the global spin-rotation $SU(2)$ is unknown, but regardless we  may expect one of two scenarios \cite{bultinck2020Ground, thomson2021Gatedefined}.
The most natural case is a ``ferromagnetic'' Hund's coupling \cite{chatterjee2020Symmetry}, which prefers: ($\nu = 0$, singlet) Each of the previously discussed phases  occurs independently in each spin-species, preserving the global spin-rotation symmetry; and ($\nu = |2|$, spin-polarized) Each of the phases occurs only in a single spin-species (with $\nu = 2$ obtained by flipping the role of particles and holes).
Because the STM signal is additive across spin-species, in both ferromagnetic cases the preceding discussion is unchanged.

On the other hand, ``anti-ferromagnetic'' (AF)
states are obtained by starting from the ferromagnetic case and applying a $\pi$-spin rotation about some axis $\hat{n}$ in only \emph{one} of the two valleys. 
For non-IVC states this rotation leaves the order (and hence the LDOS) unchanged, but for the IVC states the inter-valley coherence transforms from $\tau_{x/y} \to \tau_{x/y} (\hat{n} \cdot \mathbf{s})$ where $\hat{n}$ is 1) arbitrary for $\nu=0$, and 2) perpendicular to the spin direction for $|\nu|=2$.
This effectively shifts the IVC phase $\theta_{\textrm{IVC}}$ by $\pi$ between the two spin-species, changing our conclusions.
At both $|\nu| = 0, 2$, the AF K-IVC state is symmetric under the unitary $U_{\hat{n}} \equiv \tau_z R^\pi_{\perp}$, where $R^\pi_{\perp}$ is a $\pi$ spin-rotation about 1) any axis perpendicular to $\hat{n}$ for $\nu = 0$, or 2) the original spin axis for $|\nu|=2$. 
The \kekule-signal is odd under $U_{\hat{n}}$, enforcing an extinction.
However, unlike $\mathcal{T}'$, the perturbation $H_{\textrm{pert}} = E_B C$ is \emph{even} under $U_{\hat{n}}$, so within this approximation the \kekule-signal remains absent even for $B>0$.

Fortunately,  a magnetic field will also couple through a spin-Zeeman field, $H_{\textrm{pert}} = E_B C + E_Z s^z$. The spin-orientation of the AF K-IVC will in general lock and cant with $E_Z$, breaking $U_{\hat{n}}$ and allowing the \kekule pattern to appear.
Assessing the magnitude in this case requires a detailed knowledge of the Hund's coupling, which we thus leave to future work. 

\noindent \textit{Discussion ---}
There has already been some work on atomically-resolved STM measurement of MATBG. In particular, Ref.~\cite{jiang2019Charge}  found a nematic state at the CNP, and observed a stripe-like signal in the atomically-resolved LDOS of the AB/BA-regions. This feature is consistent with our LDOS calculations for the nSM (see \cite{Note1} for AB/BA-regions), suggesting they have identified the nSM  as the ground state of this sample. 

Theoretical and numerical work predicts that  heterostrain drives a phase transition between the K-IVC and nSM at $\nu = 0$ \cite{parker2021StrainInduced},
and between the K-IVC and IKS at $|\nu| = 2$ \cite{kwan2021Kekul}.
Since STM can measure the local heterostrain via the distortion of the Moir\`{e} lattice \cite{kerelsky2019Maximized, xie2019Spectroscopic, choi2019Electronic},
STM measurement may verify these scenarios by correlating local strain and symmetry breaking.


Finally, we note that while we have analyzed insulators of MATBG, the symmetry analysis applies more generally. Thus, it can be used to detect symmetry-breaking ordering even when the system is doped away from integer fillings. It should thus be possible to map out the stability of symmetry-breaking order with density, and correlate this order with the observed ``cascade.''
Furthermore, since magic angle twisted trilayer graphene (MATTG) features the same symmetries and band topology \cite{khalaf2019Magic, cao2021Large, hao2021Electric}, our conclusions apply to STM measurements \cite{turkel2021Twistons, kim2021Spectroscopic} of MATTG \textit{mutatis mutandis}.

\begin{acknowledgments}
We thank P. Ledwith, E. Khalaf, D. Parker, and A. Vishwanath for discussions on the role of magnetic fields in producing a K-IVC \kekule signal.
We thank N. Bultinck and S. Chatterjee for our earlier collaborations, and M. Crommie for helpful discussions.
MPZ is indebted to X. Liu, G. Farahi, CL Xiu, and A. Yazdani for a related and inspiring collaboration on the zeroth Landau-level of graphene.
JPH is supported by Princeton University Department of Physics. JPH acknowledges support from the Samsung Scholarship foundation and UC Berkeley Department of Physics during early stage of the work. TS is funded by the Masason foundation. 
MZ was supported by the ARO through the MURI program (grant number W911NF-17-1-0323) and the Alfred P Sloan Foundation.

\noindent\textit{Note added} -- During the final stage of this work, we became aware of a similar work by C\u{a}lug\u{a}ru et al. \cite{calugaru2021Spectroscopy}, which also computes the STM signal of various TBG states.

\end{acknowledgments}

\bibliography{bibliography_manuscript}

\end{document}


\begin{CJK*}{UTF8}{min}

	\title{Supplementary Material for `Detecting symmetry breaking in magic angle graphene using scanning tunneling microscopy'}
	
    \author{Jung Pyo Hong}    \thanks{These authors contributed equally}
\affiliation{Department of Physics, Princeton University, Princeton, NJ 08540, USA}
    
    \author{Tomohiro Soejima (副島智大)}
    \thanks{These authors contributed equally}
\affiliation{Department of Physics, University of California, Berkeley, CA 94720, USA}
    
    \author{M. P. Zaletel}
\affiliation{Department of Physics, University of California, Berkeley, CA 94720, USA}
\affiliation{Materials Sciences Division, Lawrence Berkeley National Laboratory, Berkeley, CA 94720, USA}
	\date{\today}
	
	\maketitle
    \tableofcontents
	
	\end{CJK*}

\section{Geometry of Moir\'{e} superlattice}

\begin{figure}[h!]
    \centering
    \includegraphics[width=0.8\linewidth]{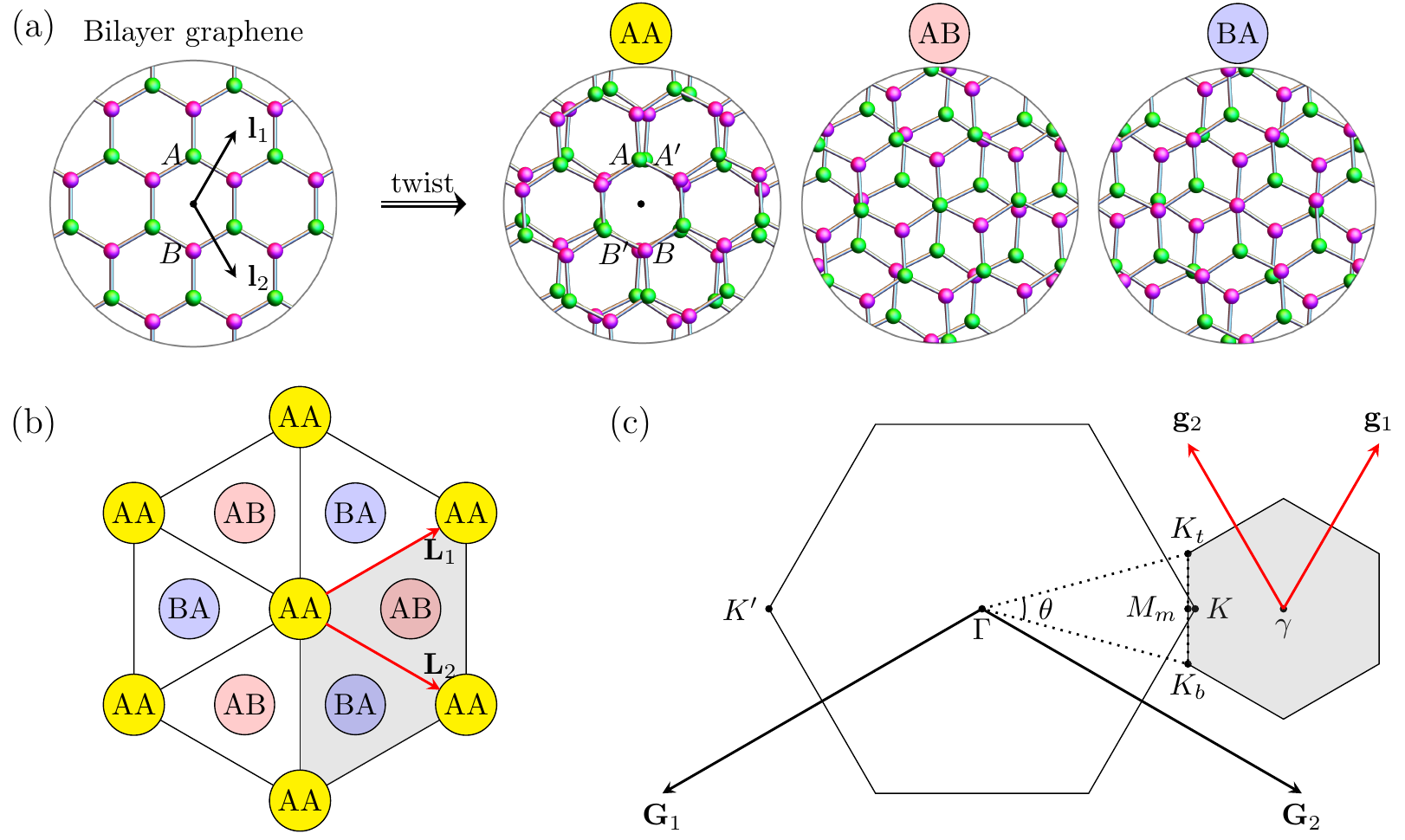}
    \caption{Geometry of a twisted bilayer graphene (TBG) lattice in real- and reciprocal- space. (See Eq.\eqref{eq:SM_lattice_geometry_graphene1}-\eqref{eq:SM_lattice_geometry_Moire3} for notations.) (a) Zoomed-in top-view of a perfectly aligned bilayer graphene (BLG) and the AA,AB,BA regions of a TBG lattice obtained after rotating by a commensurate twist angle $\theta \sim \SI{7.34}{\degree}$. Here we demonstrate $D_{6}$-configuration where rotation center (also our origin) is at the center of a plaquette, so that the TBG respects $\mathcal{C}_{2}$-symmetry. (See Fig.~\ref{fig:D3_ldos_ftldos}(a) for $D_{3}$, $D_{2}$-configurations). ($A,B$) and ($A'$,$B'$) denotes sublattice sites of the top and bottom layer, respectively. (b) Schematic illustration of the Moir\'{e} superlattice structure of TBG and its Moir\'{e} unit cell (gray region). (c) First Brillouin-Zone (BZ) of BLG and a mini-BZ (gray region) of TBG near the valley $K$.} 
    \label{fig:geometry_BZ}
\end{figure}

Geometry of a graphene lattice is defined by (Fig.~\ref{fig:geometry_BZ}(a,c))
\begin{gather} 
    \label{eq:SM_lattice_geometry_graphene1}
    \v{l}_{1} = a_{0}\left(\frac{1}{2}, \frac{\sqrt{3}}{2}\right),\,\,\,\,\,
    \v{l}_{2} = a_{0}\left(\frac{1}{2}, -\frac{\sqrt{3}}{2}\right),\,\,\,\,\, a_{0} = \SI{0.246}{\nano\meter} \\
    \label{eq:SM_lattice_geometry_graphene2}
    \v{G}_{1} = \frac{2 \pi}{a_{0}} \left(-1, \frac{1}{\sqrt{3}}\right),\,\,\,\,\,
    \v{G}_{2} = \frac{2 \pi}{a_{0}} \left(1, \frac{1}{\sqrt{3}}\right),\,\,\,\,\,
    K_{+} = K = \frac{4\pi}{3a_{0}} \left(1, 0\right) ,\,\,\,\,\,
    K_{-} = K' = \frac{4\pi}{3a_{0}} \left(-1, 0\right).
\end{gather}
Let $R(\theta)$ be a 2D rotation operator which rotates a vector by angle $\theta$ in the counter-clockwise direction. We adopt a convention where a twisted bilayer graphene (TBG) lattice is obtained by applying $R(\pm \theta/2)$ to the top and bottom graphene layer respectively. The resultant Moir\'{e} superlattice structure is defined by (Fig.~\ref{fig:geometry_BZ}(b,c))
\begin{gather} 
    \label{eq:SM_lattice_geometry_Moire1}
    \v{L}_{1}=a_{M}\left( \frac{\sqrt{3}}{2}, \frac{1}{2} \right),\,\,\,\,\, \v{L}_{2}=a_{M}\left( \frac{\sqrt{3}}{2}, -\frac{1}{2}\right),\,\,\,\,\, A_{M}=\frac{\sqrt{3}a_{M}^{2}}{2},\,\,\,\,\, a_{M}=\frac{a_{0}}{2 \sin(\theta/2)}  \\
    \label{eq:SM_lattice_geometry_Moire2}
    K_{\tau,b} = R\left( \frac{\theta}{2} \right)K_{\tau},\,\,\,\,\, K_{\tau,t} = R\left(-\frac{\theta}{2} \right)K_{\tau},\,\,\,\,\, M_{m,{\tau}}=\frac{1}{2}(K_{\tau,b} + K_{\tau,t}) \\
    \label{eq:SM_lattice_geometry_Moire3}
    \gamma_{\tau} = \frac{2\pi}{3a_{M}} \left( \tau \left( \sqrt{3} + \cot \left( \frac{\theta}{2} \right) \right), 0 \right) ,\,\,\,\,\,
    \v{g}_{1} = \frac{2\pi}{a_{M}} \left( \frac{1}{\sqrt{3}}, 1 \right),\,\,\,\,\,
    \v{g}_{2} = \frac{2\pi}{a_{M}} \left( -\frac{1}{\sqrt{3}}, 1 \right).
\end{gather}
Here $A_{M}$ denotes the area of a Moir\'{e} unit cell and $\tau=+,-$ labels the graphene valleys $K,K'$. Intervalley scattering momenta $Q_{j}$ for $j=0,\cdots,5$ are defined by
\begin{equation}
    Q_{j}= \left( R\left( \frac{\pi}{3}(j+1) \right) - R\left( \frac{\pi}{3}j \right) \right) \gamma_{+}.
\end{equation}

We used a commensurate $D_6$ lattice configuration depicted in Fig.~\ref{fig:geometry_BZ}(a) for our LDOS calculations unless otherwise specified. In the presence of strain, we followed Ref.~\cite{kwan2021Kekul,bi2019Designing} to take into of the lattice distortions.

\section{Self-consistent Hartree-Fock calculations}

We performed self-consistent Hartree-Fock calculations of the spinless interacting Bistrizer-McDonald model. We used the double-gate screened Coulomb potential $V_q = \tanh{|q| d_s}/(2|q|\epsilon_0 \epsilon_r) $ and peformed so-called subtraction procedure to avoid double counting of the Coulomb interaction. The details were reported elsewhere \cite{bultinck2020Ground, soejima2020Efficient}. We incorporated the effect of heterostrain according to Ref.~\cite{bi2019Designing}. The parameters can be found in Table.~\ref{tab:hf_params}

The IKS state was found by allowing three-fold translation breaking in $\v{L}_{1}$-direction. While some incommensurate IKS states have lower energy, commensurate IKS is believed to capture most of important physics \cite{kwan2021Kekul}.

\begin{table}[h!]
\begin{tabular}{c|c|c|c}
                                    & K-IVC             & nSM             & IKS                 \\ \hhline{=|=|=|=}
gate distance $d_s$                 & \multicolumn{3}{c}{\SI{250}{\nano\meter}}                               \\ \hline
Relative permitivity $\epsilon_r$   & \multicolumn{3}{c}{12}                                   \\ \hline
Twist angle                         & \multicolumn{3}{c}{1.05 \si{\degree}}             \\ \hline
$w_1$                               & \multicolumn{3}{c}{\SI{110}{\milli\electronvolt}}                              \\ \hline
$w_0 / w_1$                         & \multicolumn{3}{c}{0.8}                                  \\ \hline
Number of active bands                         & \multicolumn{3}{c}{2}                                  \\ \hline
Filling factor $\nu$                & \multicolumn{2}{c|}{0}              & -1                  \\ \hline
lattice dimensions $n_x \times n_y$ & \multicolumn{3}{c}{$24 \times 24 $}      \\ \hline
heterostrain $\epsilon$                         & 0 \%              & 0.04 \% & 0.2 \%           \\ \hline
\end{tabular}
\caption{Parameters used for Hartree-Fock calculation. Filling factor $\nu$ is measured relative to the charge neutrality point (CNP) in the unit of number of electrons per Moir\'{e} unit cell.}
\label{tab:hf_params}
\end{table}

\section{Spectroscopy calculations}

In this Section, we apply the Green's function formalism to self-consistent Hartree-Fock (HF) solutions of MATBG electronic system and define the spectroscopic observables (DOS,LDOS,FTLDOS) computed in this work. We further provide details on numerical evaluation procedure.

\subsection{Green's function formalism}

Let $\ket{\v{k},\tau,C}$ be the valley/Chern-polarized basis \cite{bultinck2020Ground,tarnopolsky2019Origin} for the flat bands of Bistritzer-Macdonald (BM) Hamiltonian \cite{bistritzer2011Moire}, where $\mathbf{k}$ is crystal momentum in the mini-BZ, $\tau=\pm 1$ labels the valleys $K_{\pm}=K,K'$, and $C=\pm 1$ labels the Chern number. A self-consistent eigenstate solution $\ket{\v{k},n}$ of a HF Hamiltonian $H_{\text{eff}}$ can be expressed as a linear combination of $\ket{\v{k},\tau,C}$, whose wavefunction is given by
\begin{equation} \label{eq:SM_k_n_eigenstate}
    \braket{ \mathbf{r} | \v{k},n } = \sum_{\v{g},\tau,C} F_{\tau,C}^{\v{k},n}  e^{i(\gamma_{\tau}+\v{g}+\v{k})\cdot\v{r}} \, u^{\v{k},\tau,C}_{\v{g}}(\v{r}).
\end{equation}
$F_{\tau,C}^{\v{k},n}$ are complex weights determined from the HF density matrix, $\v{g}$ are the Moir\'{e} reciprocal lattice vectors, and $u^{\v{k},\tau,C}_{\v{g}}(\v{r})$ is a cell-periodic function which satisfies $u^{\v{k},\tau,C}_{\v{g}}(\v{r})=u^{\v{k},\tau,C}_{\v{g}}(\v{r}+\v{L})$ for a commensurate Moir\'{e} superlattice.

We can recover $u^{\v{k}}_{\v{g},\tau,C}(\v{r})$ by expanding it in terms of the Wannier-orbital basis $\varphi_{\alpha}(\v{r}-\v{r}_{\alpha})$ localized at the carbon lattice sites $\v{r}_{\alpha}=\v{R}_{\alpha}+\v{t}_{\alpha}$. Here $\alpha\in(A,B,A',B')$ labels the sublattices in top layer $(A,B)$ and bottom layer $(A', B')$, while $\v{R}_{\alpha}$/$\v{t}_{\alpha}$ denotes the graphene unit cell/sublattice position vectors. Assuming $\varphi_{\alpha}(\v{r}-\v{r}_{\alpha})$ has no spatial extent in the $z$-direction for simplicity, they form an orthonormal basis set and satisfy normalization condition
$\int_{\mathbb{R}^{2}} d^{2}\v{r} \, \overline{\varphi}(\v{r}-\v{r}_{\alpha}) \varphi(\v{r}-\v{r}_{\beta}) = \delta_{\alpha,\beta}\delta_{\v{r}_{\alpha},\v{r}_{\beta}}$. 
In terms of $\varphi_{\alpha}(\v{r}-\v{r}_{\alpha})$, we can write
\begin{equation} \label{eq:SM_uk_cell_periodic}
    u^{\v{k},\tau,C}_{\v{g}}(\v{r}) =  \frac{1}{\sqrt{N_{k}N_{uc}}} \sum_{\v{R}_{\alpha},\alpha} e^{-i(\gamma_{\tau}+\v{k}+\v{g})\cdot (\v{r}-\v{R}_{\alpha}-\v{t}_{\alpha})} B^{\v{k},\tau,C}_{\v{g},\alpha} \varphi_{\alpha} (\v{r}-\v{R}_{\alpha}-\v{t}_{\alpha}).
\end{equation}
$N_{k}$ is the number of $\v{k}$-points in the mini-BZ, $N_{uc}$ is the number of graphene unit cells in a Moir\'{e} unit cell, and $B_{\v{g},\alpha}^{\v{k},\tau,C}$ are the complex coefficients obtained by diagonalizing the BM Hamiltonian.

Let $z=\omega + i\eta$ where $\omega$ is energy and $\eta>0$ is a Lorentzian broadening parameter in $\mathbb{R}$. Using Eq.\eqref{eq:SM_k_n_eigenstate}, Eq.\eqref{eq:SM_uk_cell_periodic} and spectral representation of the retarded Green's function $\hat{G}(z)=(z - H_{\text{eff}})^{-1}$ \cite{rickayzen1980Green}, total LDOS $\rho(\v{r};z)=-\Im \left( \Tr \hat{G}(z) \right)/\pi$ can be expressed as
\begin{align} \label{eq:SM_rhor_LDOS}
    \rho(\v{r};z) 
    = \frac{-1}{\pi} \Im \Bigg( \sum_{\v{k},n} \frac{\left|\braket{ \v{r} | \v{k},n }\right|^{2}}{z - E_{\v{k},n}} \Bigg)
    = \frac{-1}{\pi} \Im \Bigg( \sum_{\v{k},n,\tau,\tau',\alpha,\beta,\v{r}_{\alpha},\v{r}_{\beta}} \overline{\varphi}(\v{r}-\v{r}_{\alpha}) \frac{\Lambda^{\v{k},n}_{\tau,\tau';\alpha,\beta}(\v{r}_{\alpha},\Delta\v{r}_{\alpha\beta})}{z-E_{\v{k},n}} 
    \varphi(\v{r}-\v{r}_{\alpha}-\Delta \v{r}_{\alpha\beta}) \Bigg),
\end{align}
where 
\begin{gather} 
    \label{eq:SM_discrete_LDOS}
    \Lambda^{\v{k},n}_{\tau,\tau';\alpha,\beta}(\v{r}_{\alpha},\Delta\v{r}_{\alpha\beta}) 
    = \frac{1}{N_{k} N_{uc}}\sum_{\v{g},d\v{g}} e^{i(\v{k}+\v{g}+d\v{g}+\gamma_{\tau'}) \Delta \v{r}_{\alpha\beta}} e^{i(d\v{g}+\gamma_{\tau'}-\gamma_{\tau})\v{r}_{\alpha}} \overline{\phi}^{\v{k},n}_{\v{g},\tau,\alpha} \phi^{\v{k},n}_{\v{g}+d\v{g},\tau',\beta}. \\ \label{eq:SM_discrete_LDOS_coefficients}
    \phi^{\v{k},n}_{\v{g},\tau,\alpha} = \sum_{C} F^{\v{k},n}_{\tau,C} B^{\v{k},\tau,C}_{\v{g},\alpha},\,\,\,\,\,\,\,\,\,
    \Delta\v{r}_{\alpha\beta} = \v{r}_{\beta}-\v{r}_{\alpha}.\,\,\,\,\,\,\,\, 
\end{gather}
$E_{\v{k},n}=\braket{\v{k},n | H_{\text{eff}} | \v{k},n}$ is the HF-energy and $d\v{g}$ denotes a set of Moir\'{e} reciprocal lattice vectors chosen independently from $\v{g}$. In numerical evaluation, we truncate the sum over $\v{g}$ and $d\v{g}$ after ensuring convergence, which required $|\v{g}|<0.3/a_{0}$ and $|d\v{g}|<0.8/a_{0}$. $\Delta\v{r}_{\alpha\beta}$ is limited up to third nearest neighbors. For all LDOS calculations, we chose an $\v{r}$-meshgrid of resolution $\SI{0.01}{\nano\meter}\times\SI{0.01}{\nano\meter}$.

Total DOS of the system is then obtained by integrating $\rho(\v{r};z)$ over a Moir\'{e} unit cell $\Omega_{M}$ of area $A_{M}$, 
\begin{equation} \label{eq:SM_rhoz_DOS}
    \rho(z) = \int_{\Omega_{M}} \frac{\text{d}^{2}\v{r}}{A_{M}} \, \rho(\v{r};z).
\end{equation}

It is useful to decompose the total DOS/LDOS expressions in Eq.\eqref{eq:SM_rhor_LDOS},\eqref{eq:SM_rhoz_DOS},\eqref{eq:SM_rhoq_FTLDOS} into specific $(\tau,\tau';\alpha,\beta)$ components by using Eq.\eqref{eq:SM_discrete_LDOS}. In particular, we define `\kekule-LDOS' and `\kekule-DOS' for $\tau\neq\tau'$ as what follows:
\begin{gather} 
    \label{eq:SM_rhor_Kekule}
    \text{\kekule-LDOS :}\,\,\,\,\,\,\,\,
    \rho^{\text{\kekule}}(\v{r};z) = \frac{-1}{\pi} \Im \Bigg( \sum_{\v{k},n,\tau\neq\tau',\alpha,\beta,\v{r}_{\alpha},\v{r}_{\beta}} \overline{\varphi}(\v{r}-\v{r}_{\alpha}) \frac{\Lambda^{\v{k},n}_{\tau,\tau';\alpha,\beta}(\v{r}_{\alpha},\Delta\v{r}_{\alpha\beta})}{z-E_{\v{k},n}} 
    \varphi(\v{r}-\v{r}_{\alpha}-\Delta \v{r}_{\alpha\beta}) \Bigg) \\
    \label{eq:SM_rhoz_Kekule}
    \text{\kekule-DOS :}\,\,\,\,\,\,\,\, \rho^{\text{\kekule}}(z) = \int_{\Omega_{M}} \frac{\text{d}^{2}\v{r}}{A_{M}}\, \Big| \rho^{\text{\kekule}}(\v{r};z) \Big|.
\end{gather}
We take the absolute value in Eq.\eqref{eq:SM_rhoz_Kekule} since \kekule-LDOS can take negative value, unlike total LDOS. The \kekule-DOS serves as our proxy for the graphene-scale modulation (\kekule signal) due to intervalley scattering.

Let us now work out the analytic form of the total FTLDOS $\rho(\v{q};z)=\int_{\mathbb{R}^{2}} \text{d}^{2}\v{r} \,  e^{-i\v{q}\cdot\v{r}}\rho(\v{r};z)$. By applying continuum Fourier-transformation to Eq.\eqref{eq:SM_rhor_LDOS}, we obtain
\begin{equation} \label{eq:SM_rhoq_FTLDOS}
    \rho(\v{q};z) = \frac{-1}{2\pi i} \Bigg( \bigg( \sum_{\v{k},n}\frac{ \braket{ \v{k},n | \hat{\rho}_{\v{q}} | \v{k},n } }{z-E_{\v{k},n}} \bigg) - \overline{\bigg( \sum_{\v{k},n}\frac{ \braket{ \v{k},n | \hat{\rho}_{-\v{q}} | \v{k},n } }{z-E_{\v{k},n}} \bigg) } \Bigg) = \frac{-1}{2\pi i} \bigg( \Tr[\hat{\rho}_{\v{q}}\hat{G}(z)] - \overline{\Tr[\hat{\rho}_{-\v{q}}\hat{G}(z)]} \bigg)
\end{equation}
where $\hat{\rho}_{\v{q}}=e^{-i\v{q}\cdot \hat{\v{r}}}$ is density operator. Its matrix elements are given by
\begin{equation} \label{eq:SM_rhoq_density}
    \braket{ \v{k},n | \hat{\rho}_{\v{q}} | \v{k},n } = \int_{\mathbb{R}^{2}} \text{d}^{2}\v{r} \, e^{-i\v{q}\cdot\v{r}} \Bigg( \sum_{\tau,\tau',\alpha,\beta,\v{r}_{\alpha},\v{r}_{\beta}} \overline{\varphi}(\v{r}-\v{r}_{\alpha}) \Lambda^{\v{k},n}_{\tau,\tau';\alpha,\beta}(\v{r}_{\alpha},\Delta\v{r}_{\alpha\beta})
    \varphi(\v{r}-\v{r}_{\alpha}-\Delta \v{r}_{\alpha\beta}) \Bigg).
\end{equation}

In practice, FTLDOS data is extracted by performing a Fourier transform of experimentally measured LDOS. To best mimic this procedure, FTLDOS data presented in this work is also computed via discrete Fourier transform of numerically computed LDOS. Given LDOS signals on a meshgrid of size $[-L,L]\times[-L,L]$, we computed discrete Fourier-transform after applying a Gaussian window function $W(x,y)=\exp(-(x^2+y^2)/(2\sigma^{2}L^{2}))$ with $\sigma=0.35$. As long as the meshgrid region is sufficiently large enough to capture Moir\'{e}-scale modulations in $\rho(\v{ r};z)$ (\textit{i.e.} to resolve `internal' splitting of the FTLDOS peaks in $\mathcal{O}(\v{g})$-scale), we explicitly checked that the discrete Fourier-transformed LDOS data matches well with the exact FTLDOS obtained from the evaluation of Eq.\eqref{eq:SM_rhoq_FTLDOS},\eqref{eq:SM_rhoq_density} in momentum-space. Small spectral leakage and broadening of the Fourier modes were induced by the finite-sized window \cite{oppenheim1999discrete}, but they did not affect qualitative features near the peaks of FTLDOS and hence were negligible.

\subsection{Choice of orbitals}

Let us briefly comment on our choice of $\varphi$. We construct our phenomenological Wannier orbitals from 2D Gaussian functions localized at $\v{r}_{\alpha}$:
\begin{equation} \label{eq:SM_orbitals}
    \phi_{\alpha}(\v{r}-\v{r}_{\alpha}) = \sqrt{\frac{2}{\pi l^{2}_{orb}}} e^{-|\v{r}-\v{r}_{\alpha}|^{2}/l_{orb}^{2}},
\end{equation}
whose density $|\phi_{\alpha}(\v{r}-\v{r}_{\alpha})|^{2}$ is normalized to unity in $\mathbb{R}^{2}$. We imposed an approximate orthonormality of the orbitals by taking a superposition
\begin{equation}
    \varphi_\alpha(\v{r} - \v{r}_{\alpha}) =  \phi_{\alpha}(\v{r}-\v{r}_{\alpha}) - \sum_{\beta \in \braket{\alpha\beta}} \frac{\braket{\phi_\alpha,\phi_\beta}}{2}  \phi_{\beta}(\v{r}-\v{r}_{\beta})
\end{equation}
where the sum is over the nearest neighbor sites of $\alpha$, and $\braket{,}$ is an overlap integral between the orbitals. This ensures orthonormality up to $O(\braket{\phi_\alpha,\phi_\beta}^2)$ corrections.
    
We chose $l_{orb}\approx 0.263a_{0}$ to roughly reproduce the observed STM pattern of graphene \cite{liu2021Visualizing} in the expectation that the orbital has substantial amplitude at the nearest-neighbor bond centers. This gives a bond-centered weight of $|\phi_{\alpha}(|\v{r}-\v{r}_{\alpha}|=a_{0}/2\sqrt{3})|^{2}/|\phi_{\alpha}(0)|^{2} \approx 0.3$. In practice, $l_{orb}$ can be a fitting parameter for reproducing experiment.

\section{Derivation of \kekule-LDOS signal extinction}

We first show the case of antiunitary operator satisfying $\mathcal{K}^{-1} \hat{\rho}_\v{q} \mathcal{K} = \pm \hat{\rho}_{-\v{q}}$. Starting from Eq.\eqref{eq:SM_rhoq_FTLDOS}, we check the total FTLDOS transforms under $\mathcal{K}$ via

\begin{align} 
    \label{eq:SM_K_action_on_rhoq1}
    \rho(\v{q};z) &= \frac{-1}{2\pi i} \bigg( \Tr[\hat{\rho}_{\v{q}}\hat{G}(z)] - \overline{\Tr[\hat{\rho}_{-\v{q}}\hat{G}(z)]} \bigg)  \\
    \label{eq:SM_K_action_on_rhoq2}
    &= \frac{-1}{2\pi i} \bigg( \Tr[\mathcal{K}\mathcal{K}^{-1}\hat{\rho}_{\v{q}}\mathcal{K}\mathcal{K}^{-1}\hat{G}(z)] - \overline{\Tr[\mathcal{K}\mathcal{K}^{-1}\hat{\rho}_{-\v{q}} \mathcal{K}\mathcal{K}^{-1}\hat{G}(z)]} \bigg) \\
    \label{eq:SM_K_action_on_rhoq3}
    &= \frac{-1}{2\pi i} \bigg( \overline{\Tr[\mathcal{K}^{-1}\hat{\rho}_{\v{q}}\mathcal{K}\mathcal{K}^{-1}\hat{G}(z)\mathcal{K}]}
    -
    \Tr[\mathcal{K}\mathcal{K}^{-1}\hat{\rho}_{-\v{q}} \mathcal{K}\mathcal{K}^{-1}\hat{G}(z)\mathcal{K}] \bigg)\\
    \label{eq:SM_K_action_on_rhoq4}
    &= \frac{-1}{2\pi i} \bigg( \pm \overline{\Tr[\hat{\rho}_{-\v{q}} \hat{G}^\dagger(z)]}
    \mp
    \Tr[\hat{\rho}_{\v{q}} \hat{G}^\dagger(z)] \bigg)  = \pm \rho(\v{q};z),
\end{align}
where in going from second to third line we used skew-cyclicity of trace of antiunitary operators: $\Tr[AB] = \overline{\Tr[BA]}$ for $A$, $B$ antiunitary.

Similarly, for unitary operators satisfying $U^\dagger \hat{\rho}_\v{q} U = \pm \hat{\rho}_{\v{q}}$ we have

\begin{align} 
    \label{eq:SM_U_action_on_rhoq1}
    \rho(\v{q};z) &= \frac{-1}{2\pi i} \bigg( \Tr[\hat{\rho}_{\v{q}}\hat{G}(z)] - \overline{\Tr[\hat{\rho}_{-\v{q}}\hat{G}(z)]} \bigg)  \\
    \label{eq:SM_U_action_on_rhoq2}
    &= \frac{-1}{2\pi i} \bigg( \Tr[U^\dagger\hat{\rho}_{\v{q}}U U^\dagger \hat{G}(z)U] - \overline{\Tr[U^\dagger\hat{\rho}_{-\v{q}}U U^\dagger \hat{G}(z)U]} \bigg) \\
    \label{eq:SM_U_action_on_rhoq3}
    &= \frac{-1}{2\pi i} \bigg( \pm \Tr[\hat{\rho}_{\v{q}}\hat{G}(z)] \mp \overline{\Tr[\hat{\rho}_{-\v{q}}\hat{G}(z)]} \bigg) = \pm \rho(\v{q};z).
\end{align}

In either case, we see that the minus sign enforces $\rho(\v{q};z) =0$.

\section{Extraction of $\theta_{IVC}$}

\begin{figure}[h!]
    \centering
    \includegraphics[width=1.0\linewidth]{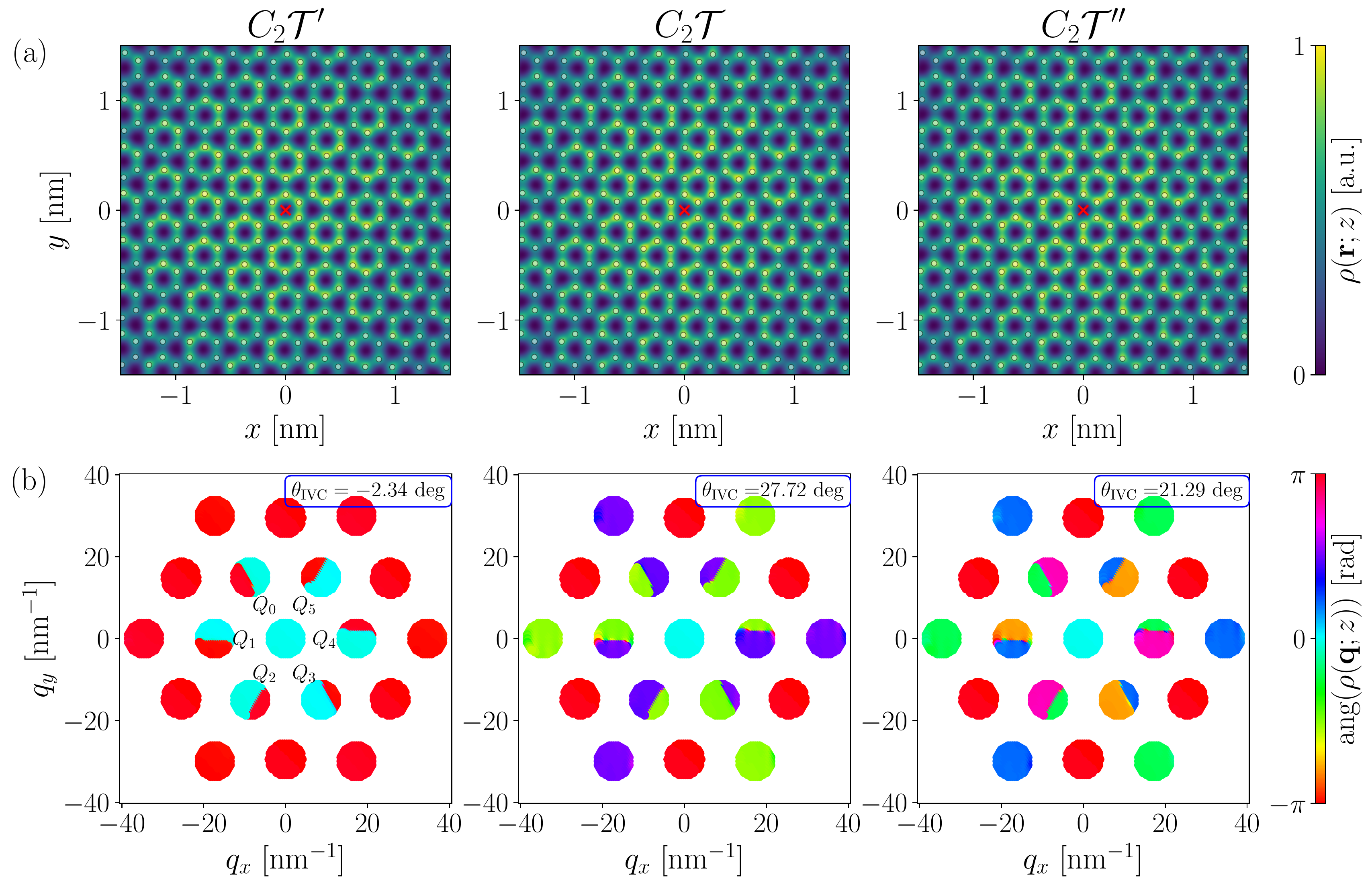}
    \caption{(a) Total LDOS $\rho(\v{r};z=\omega+i\eta)$ and (b) phases of total FTLDOS $\rho(\v{q};z)$ for the $\mathcal{C}_{2}\Tp$, $\mathcal{C}_{2}\mathcal{T}$, $\mathcal{C}_{2}\mathcal{T}''$-invariant spinless K-IVC states at CNP under magnetic field $E_{B}=\SI{0.1}{\milli\electronvolt}$, characterized by $\theta_{\text{IVC}}=\SI{0}{\degree}$, \SI{30}{\degree}, \SI{23.61}{\degree} mod \SI{60}{\degree}, respectively. Using Eq.\eqref{eq:SM_theta_IVC_from_Q0Q2Q4}, we extract $\theta_{\text{IVC}}=\SI{-2.34}{\degree}$, \SI{27.72}{\degree}, \SI{21.29}{\degree} mod \SI{60}{\degree}. The data shown is for the AA-region of bottom layer, at $\omega=\SI{-19.8}{\milli\electronvolt}$ and $\eta=\SI{0.1}{\milli\electronvolt}$.}
    \label{fig:KIVCB_IVC_ftldos_phase}
\end{figure}

In this Section, we explicitly demonstrate that the $\theta_\text{IVC}$ of K-IVC states can be extracted from LDOS measurement of K-IVC$_B$.
As explained by Liu et al.  \cite{liu2021Visualizing}, the intervalley coherent phase $\theta_{\text{IVC}}$ can be extracted from the phase structure of the FTLDOS  up to a $\pi/3$ ambiguity. 
In that work the FTLDOS was used to detect \kekule-symmetry breaking in the zeroth Landau-level of monolayer graphene in a magnetic field, and even the presence of skyrmions in the symmetry-breaking order parameter. The key ingredient was to identify particular combinations of FTLDOS signals that correspond to sublattice polarization and valley coherence.
An essentially identical procedure can be applied to MATBG, since Moir\'{e} scale modulation should not change the analysis. We review this procedure, making suitable adjustments to account for the unique symmetry properties of K-IVC$_B$.
 This suggests that K-IVC skyrmions, which is a prerequisite for the proposed skyrmionic superconductivity of MATBG \cite{khalaf2021Charged}, might be observable from STM measurements.

For simplicity we start by assuming a MATBG in $D_6$-configuration (Fig.~\ref{fig:geometry_BZ}(a)), such that the lattice has an exact $\mathcal{C}_2$-symmetry at the origin, a honeycomb plaquette center of the AA-region. Recall that a generic K-IVC state respects $\Tp$ and $\mathcal{C}_2 \mathcal{T}''$, where $\Tp=\tau_{z}\mathcal{T}$ and $\mathcal{T}''=\mathcal{T}e^{i\tau_{z}(\theta_{\text{IVC}}-\pi/2)/2}$ \cite{bultinck2020Ground}. Let us first look at `special' states, e.g. $\mathcal{C}_2 \Tp$-invariant state which must have $\theta_\text{IVC} = 0,\pi$. In the presence of a perpendicular magnetic field which breaks $\mathcal{T}$, the system still retains $\mathcal{C}_{2}$. Following an argument similar to Eq.\eqref{eq:SM_U_action_on_rhoq1}-\eqref{eq:SM_U_action_on_rhoq3}, this implies that the \kekule-FTLDOS of K-IVC$_{B}$ is purely real. In other words, \kekule-LDOS is symmetric under $\mathcal{C}_{2}$ and results in a $\mathcal{C}_{2}$-symmetric total LDOS, see Fig.~\ref{fig:KIVCB_IVC_ftldos_phase}(a). Similarly, $\mathcal{C}_{2}\mathcal{T}$-invariant states, necessarily at $\theta_{\text{IVC}}=\pm\pi/2$, preserve $\mathcal{C}_2 \tau_z$ and its \kekule-FTLDOS becomes purely imaginary (i.e. \kekule-LDOS is anti-symmetric under $\mathcal{C}_{2}$ and total LDOS loses $\mathcal{C}_{2}$-invariance, see Fig.~\ref{fig:KIVCB_IVC_ftldos_phase}(b)). Because \kekule-FTLDOS, directly sensitive to $\theta_{\text{IVC}}$, is by definition peaked at the intervalley scattering-momenta $Q_{j}$, we henceforth focus on the intervalley peaks $\rho(Q_{j})$ while assuming that the bias voltage of STM is chosen appropriately to maximize the \kekule signals.

Generic $\mathcal{C}_2 \mathcal{T}''$-invariant states with $\theta_{\text{IVC}}=\Delta\theta$ are related to the $\mathcal{C}_2\Tp$-invariant states via valley rotation $e^{i\tau_{z}\Delta \theta/2}$. This induces a phase shift $\rho(Q_{j}) \to e^{i\Delta \theta} \rho(Q_{0/2/4}), e^{-i\Delta \theta} \rho(Q_{1/3/5})$. Because the peaks are purely real for the $\mathcal{C}_2\Tp$-invariant states, we get
\begin{equation}
    \text{ang}[\rho(Q_j)] = \theta_\text{IVC} \mod \pi.
\end{equation}
In practice, we do not know the position ($r_{c}$) of the $\mathcal{C}_2$-rotation center precisely with respect to an origin of the lattice, which induces an arbitrary phase shift $\rho(Q_j) \to e^{iQ_j \cdot r_{c}} \rho(Q_j)$. To eliminate this origin-dependence, we can take a product of the peaks whose total momentum is zero. $\{Q_0, Q_2, Q_4\}$ is one such combination, so we obtain
\begin{equation} \label{eq:SM_theta_IVC_from_Q0Q2Q4}
    \text{ang}[\rho(Q_0)\rho(Q_2)\rho(Q_4)] = 3\theta_\text{IVC} \mod \pi.
\end{equation}
Thus, $\theta_\text{IVC}$ can be determined up to $\pi/3$ ambiguity. In Fig.~\ref{fig:KIVCB_IVC_ftldos_phase}, we show the total LDOS and total FTLDOS of three different K-IVC$_{B}$ states obtained via invoking valley rotations to the Hartree-Fock density matrix: (i) (approximately) $\mathcal{C}_{2}\Tp$-invariant state, (ii) $\mathcal{C}_{2}\mathcal{T}$-invariant state, and (iii) $\mathcal{C}_{2}\mathcal{T}''$-invariant state at $\Delta\theta=\SI{23.61}{\degree}$. By using Eq.\eqref{eq:SM_theta_IVC_from_Q0Q2Q4}, we extract $\theta_{\text{IVC}}=\SI{-2.34}{\degree}$, \SI{27.72}{\degree}, \SI{21.29}{\degree} mod \SI{60}{\degree} respectively, which matches with the expectations up to a small error $\sim \mathcal{O}(\SI{1}{\degree})$.

Let us comment a few noteworthy aspects about the extraction procedure. First, the FTLDOS peaks are typically broadened from the existence of Moir\'{e}-scale modulations in LDOS (e.g. due to smooth change of sublattice polarization textures), and may exhibit phase variations across nodal lines in vicinity of the peaks. In such cases, even if we choose $\rho(\v{q}=Q_{j}+\epsilon_{j})$ with small deviations $\epsilon_{j}\sim\mathcal{O}(\v{g})$, the extraction procedure will still be robust as long as $\epsilon_{0}+\epsilon_{2}+\epsilon_{4}=0$. Second,
even when the system lacks an exact $\mathcal{C}_2$-symmetry (e.g. $D_{3}$-configuration, see Fig.~\ref{fig:D3_ldos_ftldos}), the above procedure can still be used to extract $\theta_\text{IVC}$ because slowly varying part of the DOS retains an approximate $\mathcal{C}_2$-symmetry about some honeycomb plaquette near the very center of the AA-region. This allows us to leverage the origin-independence. Nevertheless for the AB/BA-regions, there exists no such $\mathcal{C}_{2}$-invariant center, so the extraction procedure above will be inapplicable. This is in contrast to the case of monolayer graphene which always has a local $\mathcal{C}_{2}$-invariant center.

We note that the local extraction of $\theta_\text{IVC}$ in different AA-regions can in principle be used to detect translation-breaking order of the IKS states. The IKS breaks translation-symmetry of the Moir\'{e} lattice but instead enjoys a ``screw''-like symmetry along the translation-broken direction (e.g. $\propto\v{L}_{1}$), which combines the translation $\v{L}_{1}$ followed by a valley rotation \cite{kwan2021Kekul}. This manifests as different \kekule patterns at different Moir\'{e} unit cells, as valley rotation takes $\theta_\text{IVC} \to \theta_\text{IVC} + 2\pi/M$, where $M \sim 3$ depends on model detail \cite{kwan2021Kekul}. However, we expect this extraction to be difficult in practice because (i) $M$ is expected to be close to $3$, which suffers from $\pi/3$ ambiguity and (ii) the Moir\'{e} lattice is not commensurate at generic strain and twist angle.

\section{Supplementary numerical data} \label{sec:SM_data}

\subsection{K-IVC$_{B}$: LDOS/FTLDOS for $D_3$-configuration} \label{sec:SM_data_D3}

\begin{figure}[h!]
    \centering
    \includegraphics[width=0.65\linewidth]{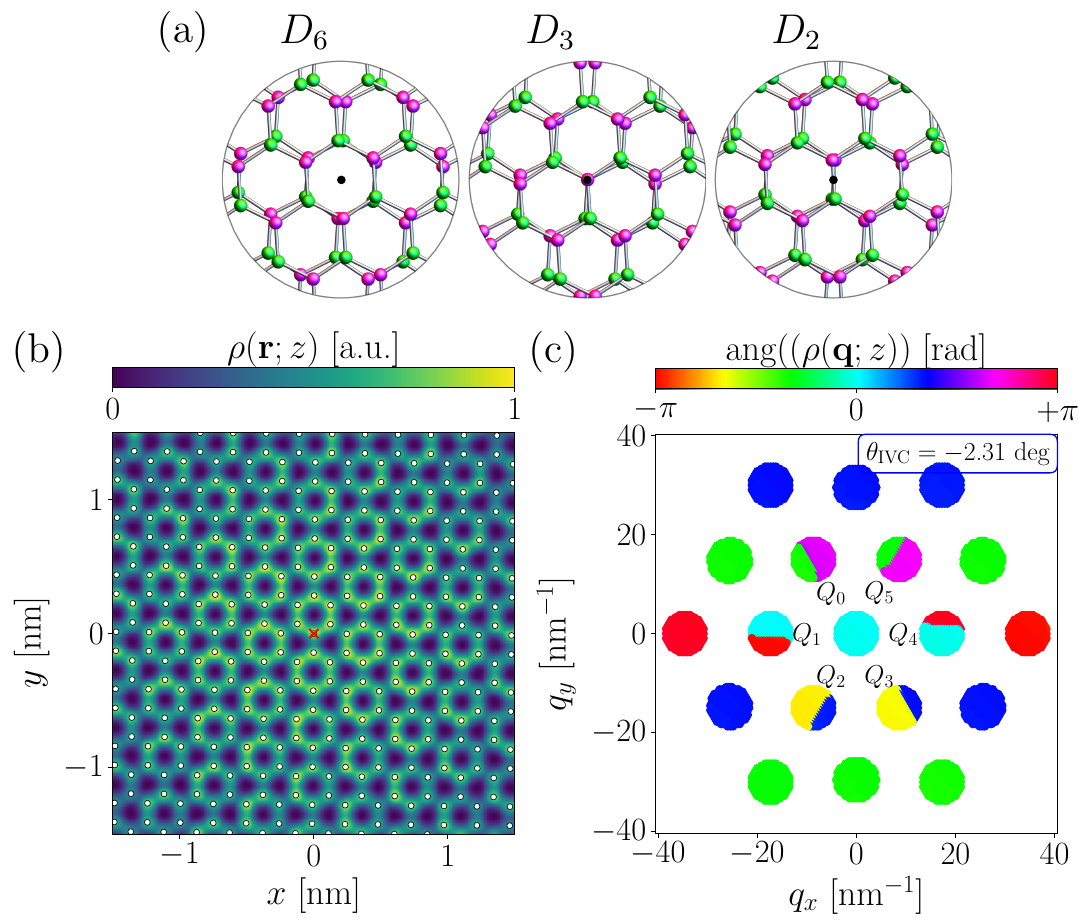}
    \caption{(a) Zoomed-in view of the AA-region of MATBG in various lattice configurations at twist angle $\theta\sim \SI{7.34}{\degree}$. Each configuration is generated by choosing a plaquette-, site-, and bond-center as the rotation-center, and respects symmetries of the point group $D_{6}$, $D_{3}$, and $D_{2}$, respectively. (b) Total LDOS $\rho(\v{r};z=\omega+i\eta)$ and (c) phases of total FTLDOS $\rho(\v{q};z)$ for the spinless, approximately $\mathcal{C}_{2}\Tp$-invariant K-IVC states at CNP under magnetic field $E_{B}=\SI{0.1}{\milli\electronvolt}$, defined on a MATBG in $D_{3}$-configuration. Using Eq.\eqref{eq:SM_theta_IVC_from_Q0Q2Q4}, we extract $\theta_{\text{IVC}}=\SI{-2.31}{\degree}$ mod \SI{60}{\degree}. The data shown is for the AA-region of bottom layer, at $\omega=\SI{-19.8}{\milli\electronvolt}$ and $\eta=\SI{0.1}{\milli\electronvolt}$.}
    \label{fig:D3_ldos_ftldos}
\end{figure}

In order to confirm that our predictions do not rely on detailed geometry (in particular the existence of an exact $\mathcal{C}_{2}$-symmetry at the AA-region) of a TBG lattice, we performed LDOS calculations on a MATBG in $D_3$-configuration (Fig.~\ref{fig:D3_ldos_ftldos}(a)). For concreteness, we computed LDOS/FTLDOS for approximately $C_{2}\Tp$-invariant K-IVC$_B$ states at $E_{B}=\SI{0.1}{\milli\electronvolt}$. As we see in Fig.~\ref{fig:D3_ldos_ftldos}(b,c), not only the \kekule-LDOS patterns are similar to those of $D_6$-configuration, the same $\theta_\text{IVC}$ can be extracted from the phases of the FTLDOS up to a small error (c.f. Fig.~\ref{fig:KIVCB_IVC_ftldos_phase}(a)).

\subsection{K-IVC: magnetic field dependence of the \kekule-DOS spectrum} \label{sec:SM_data_EBKekule}

\begin{figure}[h!]
    \centering
    \includegraphics[width=0.5\linewidth]{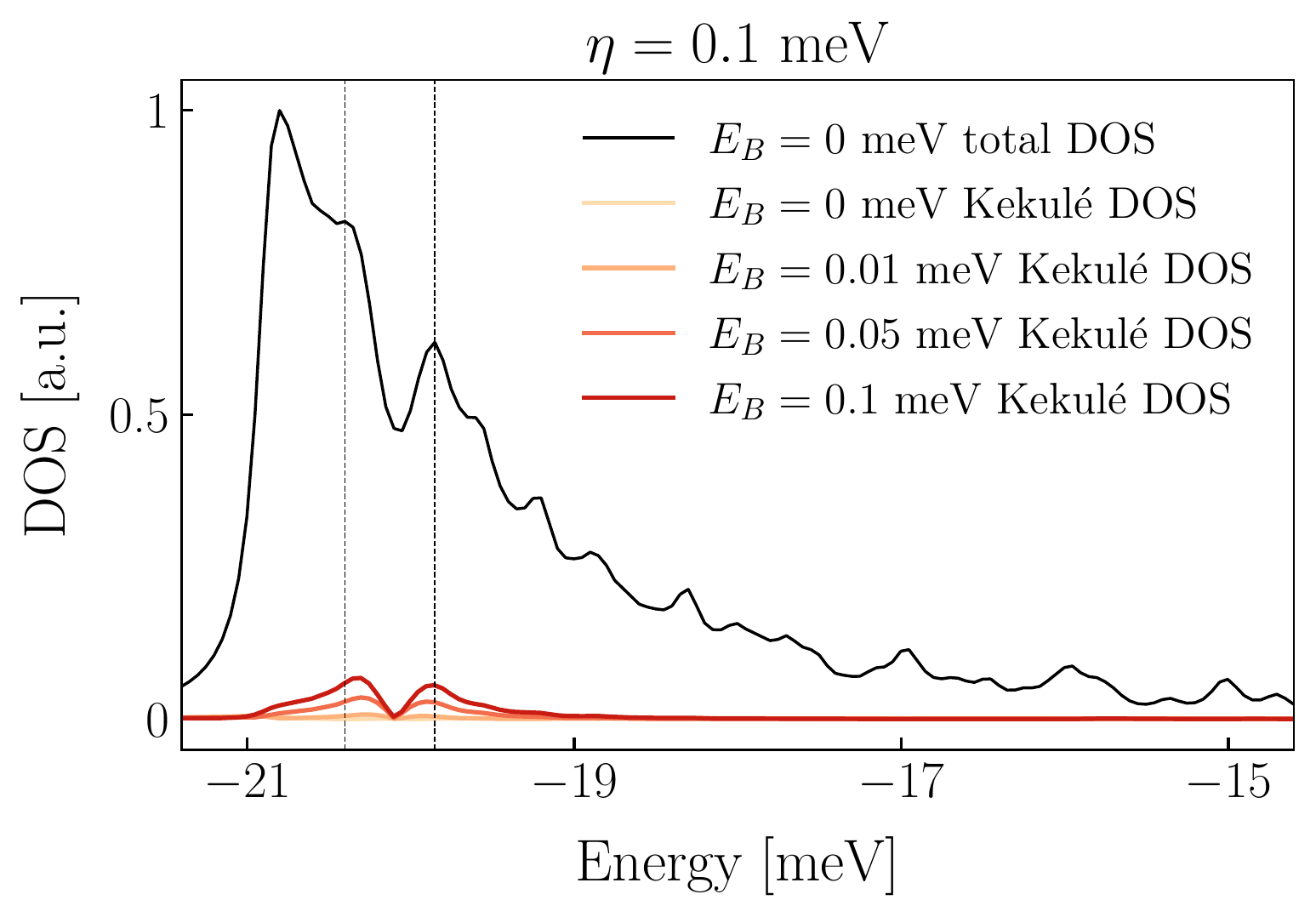}
    \caption{Total DOS $\rho(z=\omega+i\eta)$ and \kekule-DOS $\rho^{\text{\kekule}}(z)$ (Eq.\eqref{eq:SM_rhoz_Kekule}) spectrum of the valence bands for the spinless K-IVC states at CNP under various magnetic fields, at $\eta=\SI{0.1}{\milli\electronvolt}$. Dotted vertical lines identify two sub-peaks ($\sim -20.4$,$\SI{-19.9}{\milli\electronvolt}$) within a van-Hove singularity peak of the valence-bands.}
    \label{fig:EB_Kekule}
\end{figure}

We show in Fig.~\ref{fig:EB_Kekule} the total DOS and \kekule-DOS of K-IVC states under various magnetic fields at $\eta=\SI{0.1}{\milli\electronvolt}$. The \kekule-DOS signal increases approximately linearly as a function of the perturbation strength $E_{B}$, and reaches $\sim10\%$ of the total DOS signal (near the energies indicated by the dotted vertical lines) at $E_{B}=\SI{0.1}{\milli\electronvolt}$. This is consistent with our prediction in the main text that $E_B \gtrsim \eta$ is necessary for observing substantial \kekule signal. 

\subsection{nSM: comparison of the full DOS spectrum and charge-density signals at different heterestrain parameters} \label{sec:SM_nSM}

\begin{figure}[h!]
    \centering
    \includegraphics[width=1.0\linewidth]{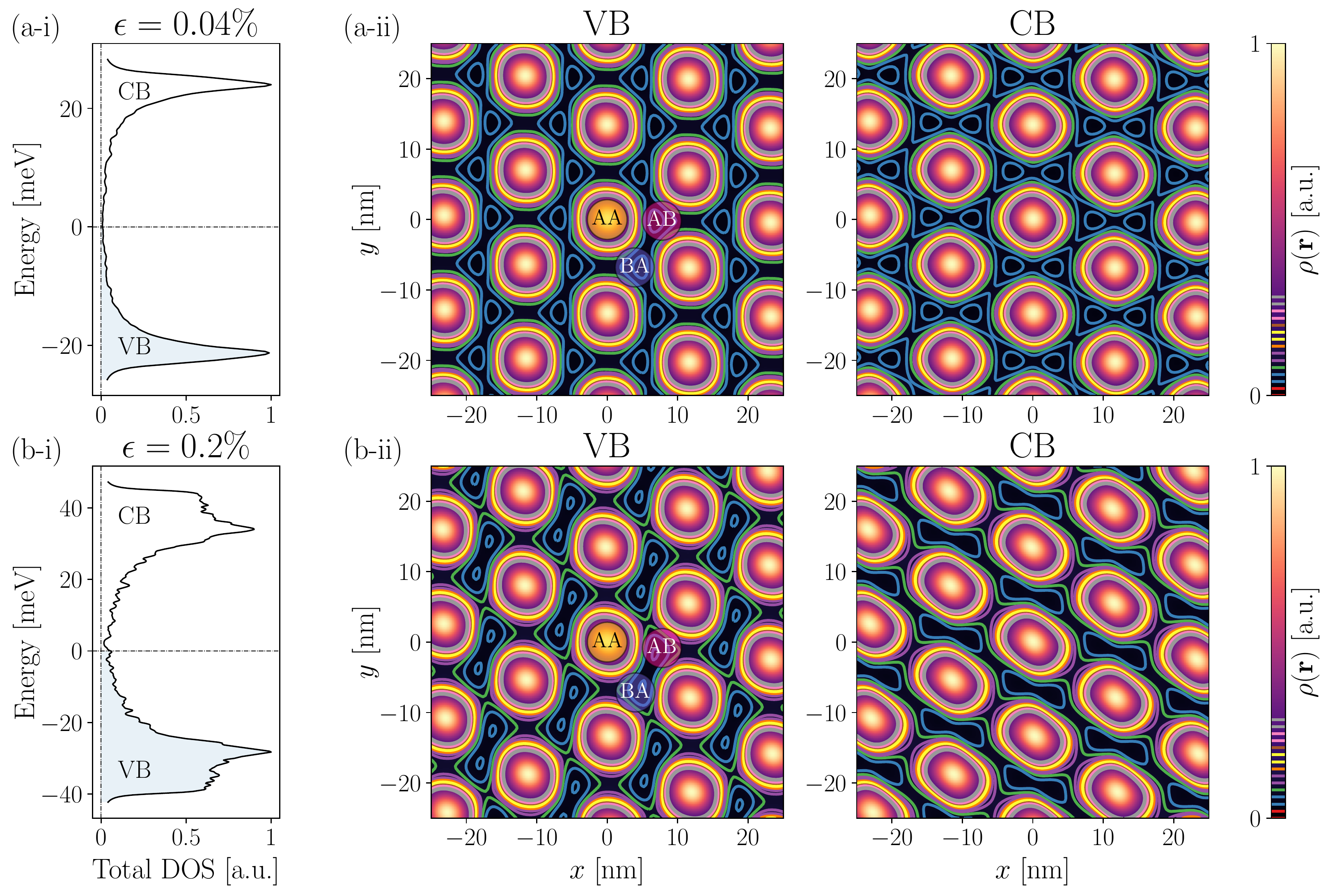}
    \caption{(i) Total DOS $\rho(z=\omega+i\eta)$ spectrum with $\eta=\SI{0.5}{\milli\electronvolt}$ and (ii) total charge-density $\rho(\v{r})$ of the conduction/valence bands (CB/VB) for the two spinless nSM states at CNP, obtained at heterostrain parameter (a) $\epsilon=0.04 \%$ and (b) $\epsilon=0.2 \%$. The electrons fill 2 bands out of the 4 bands (VB, shaded blue area of (a-i,b-i)). The colored contour lines in (a-ii,b-ii) represent the iso-signal contours within [0.0, 0.3]-range of the normalized $\rho(\v{r})$. The data shown is for the bottom layer.}
    \label{fig:nSM_cd_Moire}
\end{figure}

\begin{figure}[h!]
    \centering
    \includegraphics[width=0.75\linewidth]{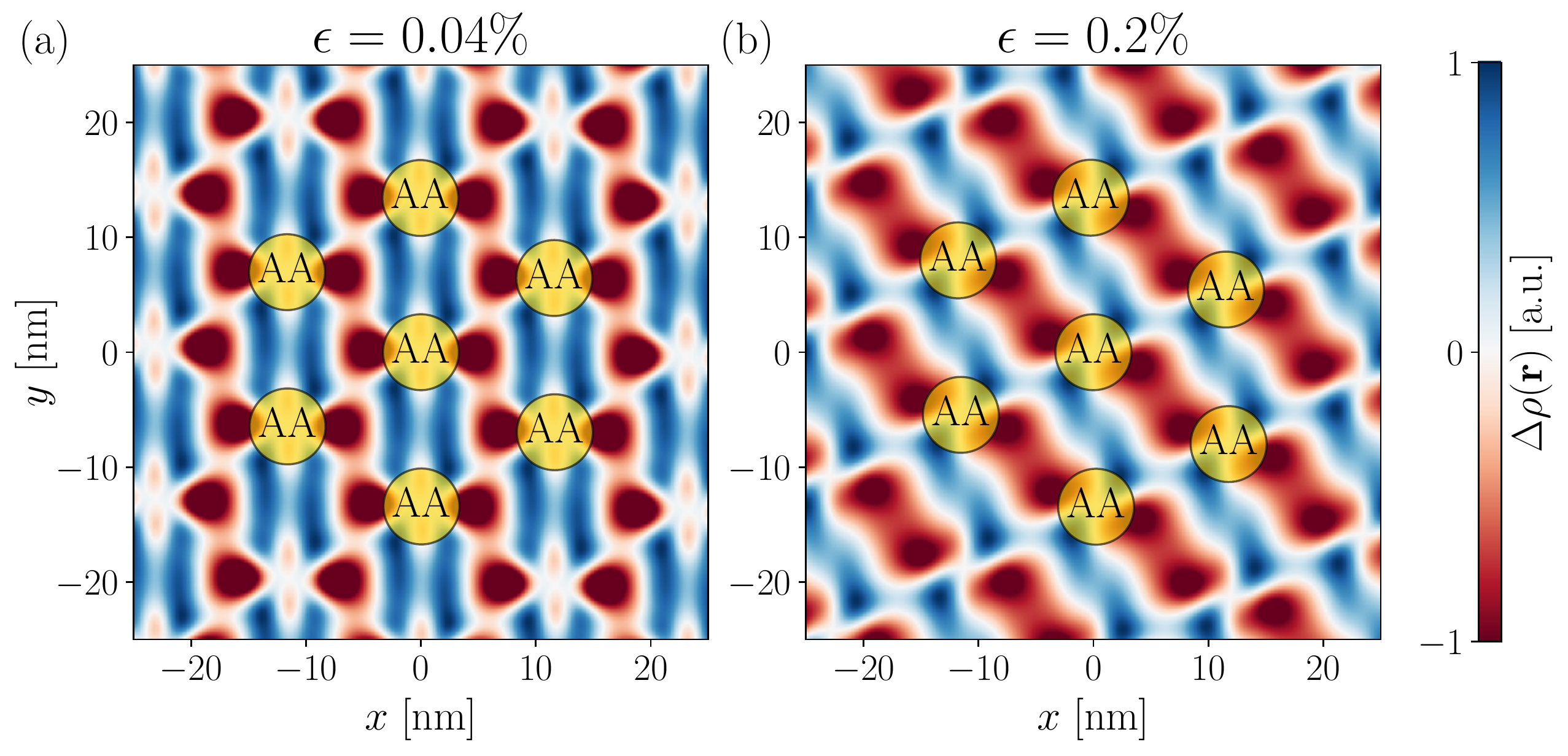}
    \caption{Net charge-density signal $\Delta\rho(\v{r})=\rho_{\text{VB}}(\v{r}) - \rho_{\text{CB}}(\v{r})$ for the nSM states shown in Fig.~\ref{fig:nSM_cd_Moire}, at heterostrain parameter (a) $\epsilon=0.04 \%$ and (b) $\epsilon=0.2 \%$. The data shown is for the bottom layer.}
    \label{fig:nSM_cd_net_charge}
\end{figure}

\begin{figure}[h!]
    \centering
    \includegraphics[width=1.0\linewidth]{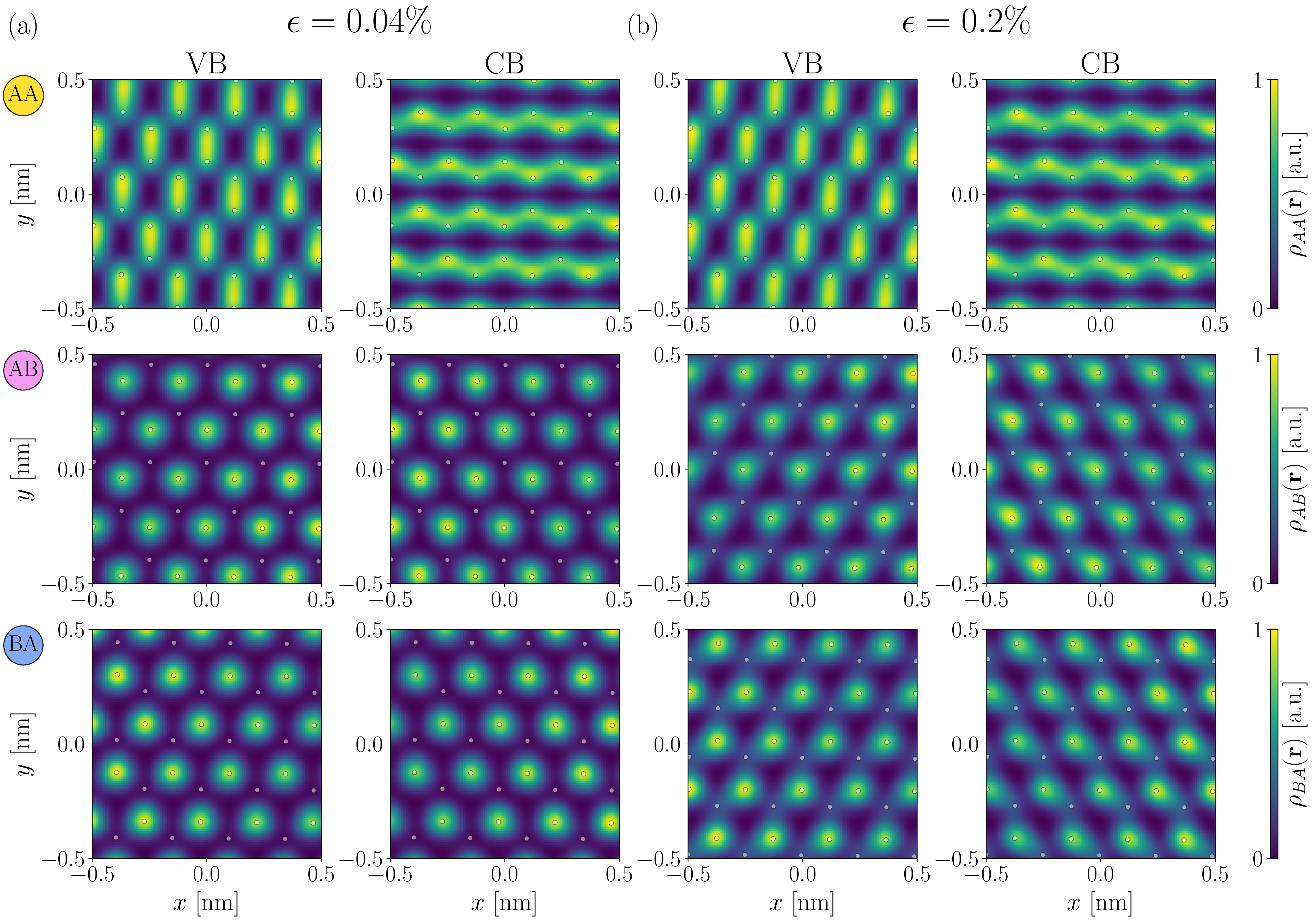}
    \caption{Graphene-scale charge-density $\rho(\v{r})$ of the conduction/valence bands (CB/VB) for the nSM states shown in Fig.~\ref{fig:nSM_cd_Moire}, at heterostrain parameter (a) $\epsilon=0.04 \%$ and (b) $\epsilon=0.2 \%$. Subplots at each row are centered at the AA/AB/BA-regions of the strained MATBG lattice, depicted in Fig.~\ref{fig:nSM_cd_Moire}(a-ii,b-ii). The data shown is for the bottom layer.}
    \label{fig:nSM_cd_AAABBA}
\end{figure}

We show in Fig.~\ref{fig:nSM_cd_Moire} the total DOS spectrum and the energy-integrated LDOS (charge-density) $\rho(\v{r})=\int \text{d}\omega \rho(\v{r};\omega)$ of the conduction/valence band (CB/VB) for the two distinct nSM states at CNP, computed at heterostrain parameters \cite{bi2019Designing} (a) $\epsilon=0.04 \%$ and (b) $\epsilon=0.2 \%$. Via shape distortion and Moir\'{e}-scale spatial distribution of the charge-density signals in the AA-regions, we can observe the global charge-nematicity and the effect of strain. Furthermore, we observe that CB$\leftrightarrow$VB switches its nematic axis; following Ref.~\cite{jiang2019Charge}, this feature can be quantified by measuring the `net-charge' $\Delta\rho(\v{r})=\rho_{\text{VB}}(\v{r}) - \rho_{\text{CB}}(\v{r})$ (Fig.~\ref{fig:nSM_cd_net_charge}), where the blue/red signifies the degree of electron/hole doping. We see clearly the emergence of a stripy, quadrupolar net-charge order at the AA-regions (more prominently for $\epsilon=0.2\%$), which qualitatively matches with an experimental data \cite{jiang2019Charge}. 

We show in Fig.~\ref{fig:nSM_cd_AAABBA} the graphene-scale charge-density of the CB/VB for the same nSM states shown in Fig.~\ref{fig:nSM_cd_Moire} at (a) $\epsilon=0.04\%$ and (b) $\epsilon=0.2\%$, focusing on the very center of the AA/AB/BA-regions in bottom layer. From the AA-region, we see that the direction of local bond-anistropy switches as VB$\leftrightarrow$CB for both strain values, in a similar spirit to the switch of nematicity observed in Moir\'{e}-scale signals. At the AB/BA-regions the signals are dominantly sublattice $B$/$A$-polarized, same for the CB/VB. This can be explained from the chiral limit phenomenology of the valley/Chern basis $\ket{\v{k},\tau,C}$ \cite{bultinck2020Ground,tarnopolsky2019Origin} for the flat bands which are $\sigma=A/B=C\tau$-polarized; as shown by Ref.~\cite{tarnopolsky2019Origin}, each $A/B$-polarized flat band wavefunction then has zero density at the AB/BA-region respectively, resulting in local sublattice polarization. This phenomenology survives away from the chiral limit, as evidenced by the polarization of CB$+$VB. 

Comparing the strain-dependence of the AB/BA-region signals, we see a more striking feature, the prominence of $B$/$A$-centered ``tadpole-like'' bond-anistropy along a zigzag direction (\SI{\pm60}{\degree} from the $y$-axis) for $\epsilon=0.2 \%$, in contrast to the strong equal-weight bond-anistropy observed in the AA-region for both heterostrain values. A similar qualitative feature has been observed in experiment as well \cite{jiang2019Charge}. Such stripy bond-nematicity in the AB/BA-regions (with graphene translation symmetry) could be a signature of a non-IVC phase stabilized under a weak strain. Along with the quadrupolar net-charge distributions shown in Fig.~\ref{fig:nSM_cd_net_charge}, our observations suggest that strain significantly alters the charge order of an nSM state.

We comment that the observed bond-nematicity patterns should be sensitive to choice of the Wannier orbitals and direction of the applied strain, which can enter as tuning parameters for reproducing experiment.

\subsection{IKS: full DOS spectrum and charge-density signals at the AA/AB/BA-regions} \label{sec:SM_data_IKS}

\begin{figure}[h!]
    \centering
    \includegraphics[width=0.5\linewidth]{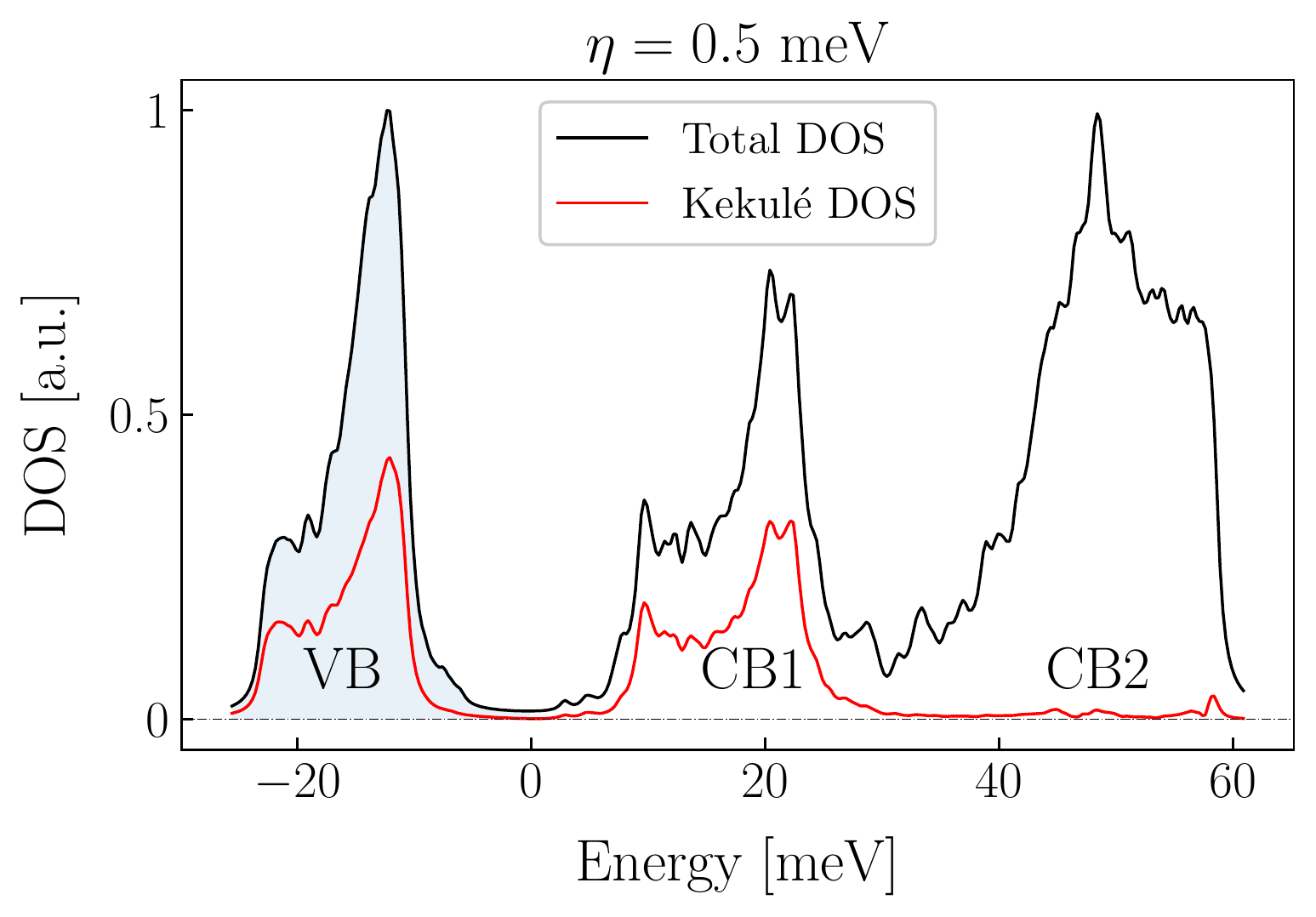}
    \caption{Total DOS $\rho(z=\omega+i\eta)$ and \kekule-DOS $\rho^{\text{\kekule}}(z)$ (Eq.\eqref{eq:SM_rhoz_Kekule}) spectrum of the spinless IKS states at filling $\nu=-1$, $\eta=\SI{0.5}{\milli\electronvolt}$, and heterostrain $\epsilon=0.2 \%$. The electrons fill 3 valence bands (VB, shaded blue area) out of the 12 bands to form an insulating state. The conduction bands are grouped into CB1 and CB2 separated by a dip near $\SI{30}{\milli\electronvolt}$, corresponding to holes filling 3 and 6 bands respectively.}
    \label{fig:IKS_dos}
\end{figure}

\begin{figure}[h!]
    \centering
    \includegraphics[width=0.75\linewidth]{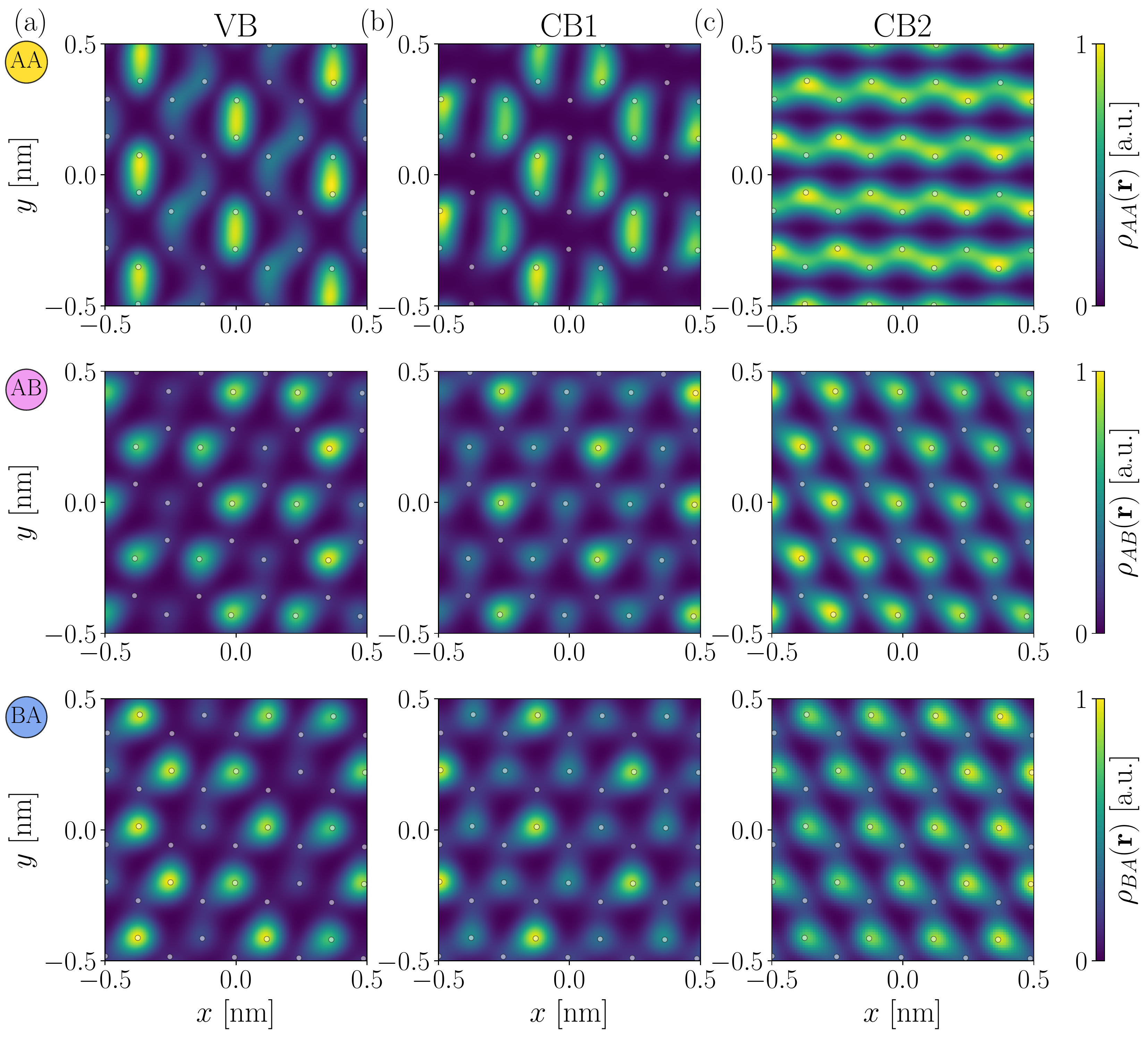}
    \caption{Graphene-scale charge-density $\rho(\v{r})$ of the valence/conduction bands (a) VB (b) CB1 (c) CB2 for the IKS states shown in Fig.~\ref{fig:IKS_dos}, at filling $\nu=-1$ and heterostrain $\epsilon=0.2 \%$. Subplots at each row are centered at the AA/AB/BA-regions of the strained MATBG lattice (c.f. Fig.~\ref{fig:nSM_cd_Moire}(a-ii,b-ii)). The data shown is for the bottom layer.}
    \label{fig:IKS_cd_AAABBA}
\end{figure}

We show in Fig.~\ref{fig:IKS_dos} the total DOS and \kekule-DOS spectrum of the spinless IKS states at $\nu=-1$ and heterostrain $\epsilon=0.2 \%$. There are 12 bands total available due to three-fold translation symmetry breaking along $\v{L}_{1}$ \cite{kwan2021Kekul}. The IKS is an insulating state at $\nu=-1$ with a band gap $10-\SI{20}{\milli\electronvolt}$, where the electrons fill 3 out of the 12 bands (VB). In contrast to the VB and low-energy conduction bands (CB1) which have a sizable \kekule-DOS, we find that the \kekule-DOS signal is significantly reduced for the higher-energy conduction bands (CB2) at $\omega \gtrsim \SI{30}{\milli\electronvolt}$. Because there exists no symmetries such as $\Tp$ in IKS which enforce selection rules in $\rho_{\v{q}}$ (Eq.\eqref{eq:SM_K_action_on_rhoq1},\eqref{eq:SM_U_action_on_rhoq1}), the absence of \kekule-DOS implies that the CB2 has lost intervalley coherence. Such ``IVC-depletion'' of the IKS has been theoretically analyzed via energetic considerations of the momentum-dependent valley-hybridization pattern, and is consistent with high valley-polarization predicted for the upper bands (see Ref.~\cite{kwan2021Kekul} for detail.)

We show in Fig.~\ref{fig:IKS_cd_AAABBA} the graphene-scale charge-density of the valence/conduction bands (VB, CB1, CB2) for the same IKS states at the very center of the AA/AB/BA-regions. For CB2, we find that the \kekule pattern disappears as expected from the vanishing \kekule-DOS while the signal resembles those of the nSM conduction bands (c.f. Fig.~\ref{fig:nSM_cd_AAABBA}), indicating a non-IVC but nematic charge order. For VB and CB1, we see clear $\sqrt{3}\times\sqrt{3}$ \kekule patterns as well as sublattice polarizations in the AB/BA-regions. The equal-weight bond-anisotropy in the AA-region and the stripy ``tadpole-like'' bond-anisotropy in the AB/BA-regions are also observed (c.f. Fig.~\ref{fig:nSM_cd_AAABBA}). Such resemblances to the nSM charge order under the same heterostrain $\epsilon=0.2 \%$ shows that a strain leaves sharp density fingerprints on IVC states as well, via bond-anisotropic charge order.



\FloatBarrier

\bibliography{bibliography_manuscript, bibliography_SM}